\documentclass[twocolumn]{aastex631}
\usepackage{amsmath}
\shorttitle{Neutrino Multiplet}
\shortauthors{Yoshida et al.}
\graphicspath{{./}{figures/}}

\begin{document}

\title{Identifying High Energy Neutrino Transients by Neutrino Multiplet-Triggered Followups}

\author[0000-0003-2480-5105]{Shigeru Yoshida}
\affiliation{International Center for Hadron Astrophysics, Chiba University,
Chiba 263-8522, Japan}

\author[0000-0002-5358-5642]{Kohta Murase}
\affiliation{Department of Physics; Department of Astronomy \& Astrophysics; Center for Multimessenger Astrophysics,  Institute for Gravitation and the Cosmos,
The Pennsylvania State University, University Park, PA 16802, USA}
\affiliation{Center for Gravitational Physics, Yukawa Institute for Theoretical Physics, Kyoto University, Kyoto, Kyoto 606-8502, Japan}

\author[0000-0001-8253-6850]{Masaomi Tanaka}
\affiliation{Astronomical Institute, Tohoku University, Sendai 980-8578, Japan}
\affiliation{Division for the Establishment of Frontier Sciences, Organization for Advanced Studies, Tohoku University, Sendai 980-8577, Japan}

\author[0000-0001-6857-1772]{Nobuhiro Shimizu}
\affiliation{International Center for Hadron Astrophysics, Chiba University,
Chiba 263-8522, Japan}

\author[0000-0003-3467-1956]{Aya Ishihara}
\affiliation{International Center for Hadron Astrophysics, Chiba University,
Chiba 263-8522, Japan}

\begin{abstract}
Transient sources such as supernovae (SNe) and tidal disruption events are candidates of high energy neutrino sources. However, SNe commonly occur in the universe 
and a chance coincidence of their detection with a neutrino signal cannot be avoided, 
which may lead to a challenge of claiming their association with neutrino emission.
In order to overcome this difficulty, we propose a search for $\sim10-100$~TeV neutrino multiple events within a timescale of $\sim 30$~days coming from the same direction, called neutrino multiplets.
We show that demanding multiplet detection by a $\sim 1$~km$^3$ neutrino telescope limits distances of detectable neutrino sources,
which enables us to identify source counterparts by multiwavelength observations
owing to the  substantially reduced rate of the chance coincidence detection of transients. 
We apply our results by constructing a feasible strategy for optical followup observations and demonstrate that wide-field optical telescopes with a $\gtrsim4$~m dish should be capable of identifying a transient associated with a neutrino multiplet. 
We also present the resultant sensitivity of multiplet neutrino detection as a function of the released energy of neutrinos and burst rate density. 
A model of neutrino transient sources with an emission energy greater than ${\rm a~few}\times 10^{51}$~erg and a burst rate rarer than ${\rm a~few}\times 10^{-8}\ {\rm Mpc}^{-3}\ {\rm yr}^{-1}$ is constrained by the null detection of multiplets by a $\sim 1$~km$^3$ scale neutrino telescope. 
This already disfavors the canonical high-luminosity gamma ray bursts and jetted tidal disruption events as major sources in the TeV-energy neutrino sky.
\end{abstract}



\section{Introduction} \label{sec:intro}


High energy neutrino astronomy has rapidly grown in recent years. The discovery of high energy cosmic neutrinos (\citealt{Aartsen:2013bka,Aartsen:2013jdh,Aartsen:2015rwa}) by the IceCube Neutrino Observatory (\citealt{Aartsen:2016nxy}) initiated neutrino sky observations to quantify the flux of the high-energy cosmic neutrino background radiation (\citealt{Aartsen:2016xlq,Aartsen:2020aqd,Aartsen:2018vtx}), measure of the neutrino flavor ratio (\citealt{Aartsen:2015ivb,Aartsen:2018vez}), and provide hints of the breakdown into a set of individual astronomical object radiating neutrinos \citep{Aartsen:2019fau}. 
Furthermore, the possible association of the high energy neutrino signal, IceCube-170922A, with the high energy gamma-ray flare detected by Fermi-LAT suggests that the blazar TXS0506+056 is a high energy neutrino source \citep{IceCube:2018dnn}, which, if this is indeed true, is the first identification of a high energy neutrino emitter. This has proven the power of {\it multimessenger} observations. Multiwavelength observation campaigns prompted by high energy neutrino detection alerts may lead to discovery of yet unknown transient neutrino sources. These sources are integral in studying the origin of high energy cosmic rays which has so far proved difficult.cosmic rays which has so far proved difficult.
We note that multiwavelength follow-up of astrophysical neutrino candidates is fundamentally complimentary to neutrino follow-ups of EM transients. For the latter, analysts must pre-select certain classes of objects for follow-up, e.g. gamma-ray bursts. This class of objects may or may not be true neutrino emitters, and therefore the analysis is limited in its discovery power. By contrast, astrophysical neutrinos are guaranteed to have specific (if faint) sources, and so multiwavelength follow-ups can remain source class agnostic, and open to discovering potentially unexpected source classes.

Conducting followup multimessenger observations triggered by a neutrino detection, to find the association with a rare class of object, is a straightforward process, since the chance of finding a transient source in the direction of the detected neutrinos that is not the emitter, referred to as {\it contamination} in the literature, is substantially suppressed.
Gamma-ray blazars belong to this category of objects which is 
why the observational indications of the possible associations 
of the blazars with neutrino emissions began to emerge.
However, the majority of high energy neutrino sources are not gamma-ray blazars \citep{Aartsen:2016lir,Murase:2016gly}. 
Another class of gamma-ray bright sources, Gamma-ray bursts (GRB)~\citep{Waxman:1997ti}, also cannot be responsible for the bulk of the all-sky neutrino flux measured by IceCube~\citep{Aartsen:2016qcr}.
Rather, research has suggested that more abundant classes of objects may be a major source, especially in the 10-100~TeV range. They include transient sources such as core-collapse supernovae (CC SNe) \citep{Murase:2010cu,Katz:2011zx,Murase:2017pfe}, low-luminosity Gamma-ray bursts (LL GRB) and trans-relativistic SNe \citep{Murase:2006mm,Gupta:2006jm,Kashiyama:2013qet}, 
jet-driven SNe~\citep{Meszaros:2001ms,Murase:2013ffa,Senno:2015tsn,Denton:2017jwk},
wind-driven transients~\citep{Murase:2009pg,Fang:2018hjp,Fang:2020bkm}, 
and (non-jetted) tidal disruption events (TDE)
\citep{Hayasaki:2019kjy,Murase:2020lnu,Winter:2022fpf}.
As many of these are known as optical transient events, an optical/NIR followup observation could find the neutrino associated transient~\citep{Murase:2006mm,Kowalski:2007xb}. However, the larger populations cause significant contamination in the optical followups. 
For example, $\sim$100 SNe are found up to a redshift $z\lesssim 2$ within 1 deg$^2$ for a duration of a few days to months, which is a typical timescale for neutrino emission from SNe, and which makes it challenging to claim robust associations between a neutrino detection and its optical counterpart candidate. 

\begin{figure}[tb!]
\plotone{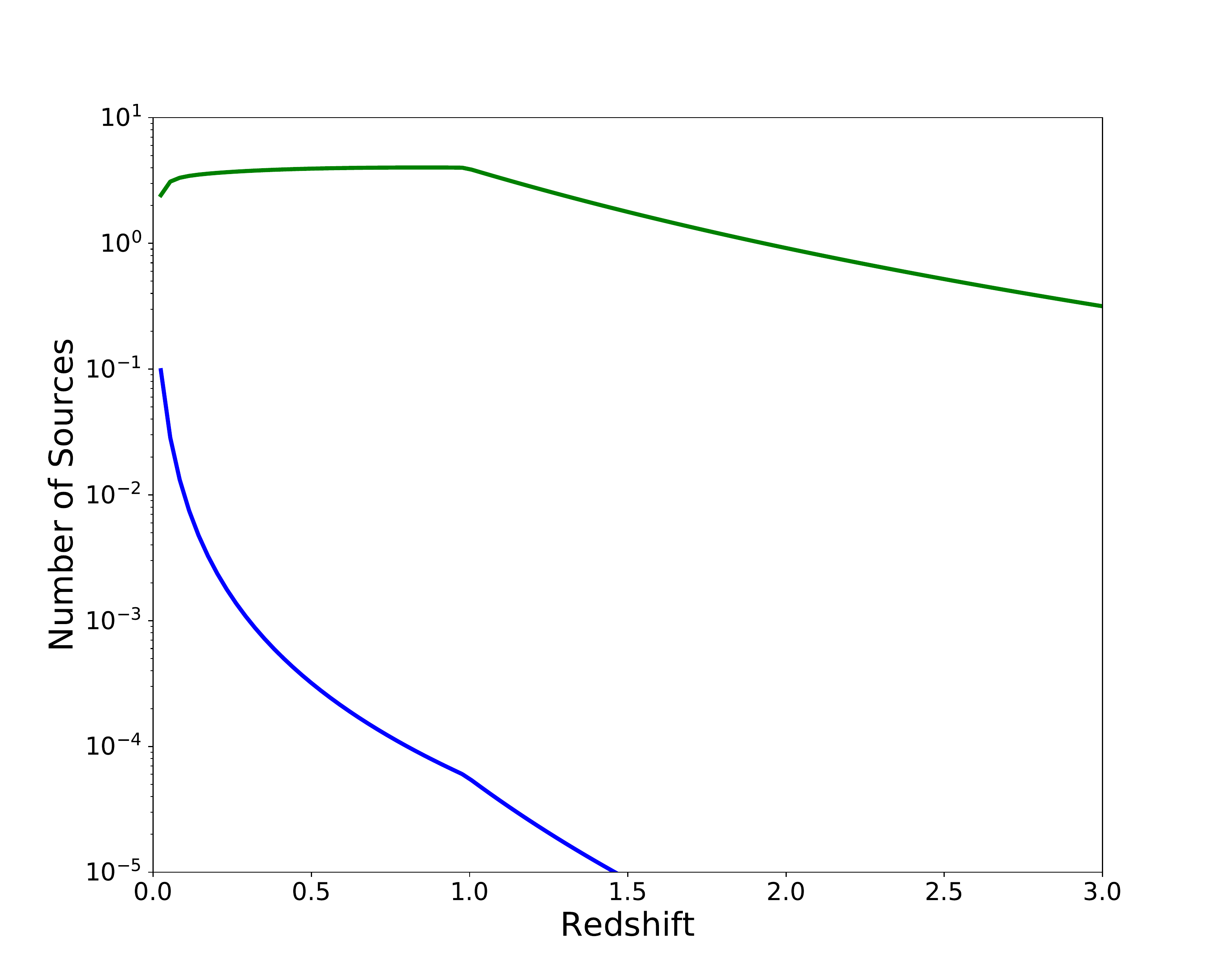}
\caption{Number of neutrino sources per redshift bin width 
$\Delta z=0.03$ in the 2$\pi$ sky to produce a singlet event (green) and multiplet events (blue).
A case of the released energy of neutrino emission ${\mathcal E}_\nu^{\rm fl}=3\times 10^{49}\ {\rm erg}$, 
the burst rate $R_0 = 3\times 10^{-6}\ {\rm Mpc^{-3} yr^{-1}}$, the flare duration of ${\Delta T}=30$~days
is presented for illustrative purpose. 
The cosmic evolution tracing the star-formation rate (SFR) 
is assumed in this model.
\label{fig:NSources4intro}}
\end{figure}

A possible solution to overcome this is to search for neutrino multiplets, two (doublet) or more neutrinos originating from the same direction within a certain time frame.
Only sources in the neighborhood of our galaxy can have 
an apparent neutrino emission luminosity
high enough to cause the detection of a neutrino multiplet given the sensitivity of the current and future neutrino telescopes. This is analogous to how, in optical astronomy,  a smaller dish telescope is only sensitive to a brighter magnitude, and thus automatically limits the distance of the observable objects for a given luminosity.
Figure~\ref{fig:NSources4intro} shows the redshift distribution of neutrino sources with a neutrino emission energy of ${\mathcal E}_\nu^{\rm fl}=3\times 10^{49}$~erg yielding singlet and multiplet neutrino detections by a 1 km$^3$ neutrino telescope. The distribution of sources to produce a singlet neutrino detection extends up to $z\gtrsim 2$ while those responsible for the multiplet neutrinos are localized. Distant transient sources cannot be associated with the neutrino multiplet, and thus a followup observation would be less contaminated by unrelated transients if measurement of the distance (or redshift) to each of the transient sources is available.

As the atmospheric neutrino background dominates the detections of high energy cosmic neutrinos, requiring multiple neutrino detections for followup observations is beneficial. Burst-like neutrino emissions, expected to be generated by, for example, prompt emissions from internal shocks in the jets of GRBs~\citep{Waxman:1997ti}, allows for the search of emitters to be restricted to tens of seconds, removing any possible contamination from background neutrinos.
\citet{Aartsen:2017snx} used this to search for neutrino multiplets from short transients.
However, many models of high-energy neutrino emissions associated with optical transients predict a longer duration. We expect neutrino flares within timescales of days to months for CC SNe (including engine-driven SNe) and TDEs.
While increasing the observational time windows significantly worsens the signal-to-noise ratio of the search, requiring neutrino doublet detections improves the ratio as, when the expected number of atmospheric neutrinos $\mu_{\rm atm}$ is less than one, the Poisson probability of recording a doublet is $\textstyle{\sim \mu_{\rm atm}^2/2}$.

In this study, we investigate the strategy of obtaining multimessenger observations
by searching for high energy multiple neutrino events, considering a 1 km$^3$ neutrino telescope like the IceCube Neutrino Observatory, and the expected sensitivity in the parameter space to neutrino transient sources.
We conduct a case study with a search time window of $T_{\rm w}=30$~days given many neutrino emission events can be characterized by this timescale.
We construct a generic model of emitting neutrino sources with energies of $\varepsilon_0\simeq$
100 TeV and show the number of sources expected to yield the neutrino multiplet. 
Further, we discuss the sensitivity to neutrino sources given changes in
source parameters, such as luminosity, considering the limitations imposed by the atmospheric neutrino background.
We propose an optical followup observation scheme to
filter out the contaminating sources and identify the object responsible
for the neutrino multiplets. Finally, we discuss the implications
to the neutrino source emission models.

A standard cosmology model with $H_0=73.5$ km $\sec^{-1}$ Mpc$^{-1}$,
$\Omega_M = 0.3$, and $\Omega_{\Lambda}=0.7$ is assumed throughout the paper.

\section{Neutrino multiplet detection} \label{sec:multiplet}

\subsection{Generic model of neutrino sources that yield neutrino multiplet detection} \label{subsec:multipletModel}


The Emission of neutrinos from transient sources can be characterized by the integral luminosity $L_\nu$ (defined for the sum of all flavors), the flare duration $\Delta T$ in the source frame, 
and the neutrino energy spectrum $\phi_\nu^{\rm fl}$.
The total energy output by a neutrino emission is given by
${\mathcal E}_\nu^{\rm fl}=L_\nu\Delta T$.

The neutrino spectrum $\textstyle{{d\dot{N}_{\nu_e+\nu_\mu+\nu_\tau}}/{d\varepsilon_\nu}}$ 
is assumed to follow a power-law form,
with the reference energy $\varepsilon_0$ 
and the flux from a single source at the redshift $z$ is given
as
\begin{eqnarray}
  \phi_\nu^{\rm fl} &\equiv& \frac{1}{4\pi d_z^2}\frac{d\dot{N}_{\nu_e+\nu_\mu+\nu_\tau}}{d\varepsilon_\nu}\nonumber\\ 
  &= & \frac{1}{4\pi d_z^2}\frac{\kappa}{\varepsilon_0} \left(\frac{\varepsilon_\nu}{\varepsilon_0}\right)^{-\alpha_\nu} \nonumber \\
  &= & \frac{1}{4\pi d_z^2}\frac{\kappa}{\varepsilon_0} \left(\frac{E_\nu(1+z)}{\varepsilon_0}\right)^{-\alpha_\nu} ,
  \label{eq:point_source_flux}
\end{eqnarray}
where $\varepsilon_\nu$ and 
$E_\nu=\varepsilon_\nu{(1+z)}^{-1}$ are the neutrino energies at the time of emission and arrival at Earth’s surface, respectively.
In our model, the normalization constant $\kappa$ is 
bolometrically associated with the source luminosity as
\begin{equation}
 \kappa = \frac{L_\nu}{\int\limits_{0.1\varepsilon_0}^{10\varepsilon_0} d\varepsilon_\nu\left(\frac{\varepsilon_\nu}{\varepsilon_0}\right)^{-\alpha_\nu+1}},
 \label{eq:luminosity_normalization}
\end{equation}
by integrating over the energy range $[0.1\varepsilon_0, 10\varepsilon_0]$
to account the neutrino energetics around $\varepsilon_0$ reasonably.
The proper distance, $d_z$, is calculated via
\begin{equation}
d_z = \frac{c}{H_0}\int\limits_0^{z} dz^{'} 
\frac{1}{\sqrt{\Omega_{\rm M}(1+z^{'})^3+\Omega_\Lambda}}.
  \label{eq:proper_distance_z}
\end{equation}

We hereafter adopt 
$\alpha_\nu=2.3$, $\varepsilon_0=100\ {\rm TeV}$, and $\Delta T= 30\ {\rm days}$ in our generic model
construction. These are set to be 
consistent with current neutrino data ($\alpha_\nu$),
within the representative energy range 
in IceCube measurements ($\varepsilon_0$)~\citep{IceCube:2021uhz}, and within the expected timescale
of neutrino flares in the majority of optical transient sources ($\Delta T$).

A population of neutrino sources contribute to the diffuse cosmic background flux.
Assuming emission from standard candles (i.e., identical
sources over redshifts),
the energy flux of diffuse neutrinos from these sources across the universe,
$\Phi_\nu\equiv dJ_{\nu}/dE_{\nu}$, is calculated by (e.g.,~\citealt{Murase:2015xka})

\begin{multline}
E_\nu^2\Phi_\nu(E_\nu)=\frac{c}{4 \pi}
\int\limits_{z_{\rm min}}^{z_{\rm max}}\frac{dz}{1+z} \left|\frac{dt}{dz}\right| \\
\times \left[\varepsilon_\nu^2\frac{d\dot{N}_{{\nu_e+\nu_\mu+\nu_\tau}}}{d\varepsilon_\nu}(\varepsilon_\nu)\right]n_0\psi(z), 
\label{eq:diffuse_flux}
\end{multline}
where $n_0\psi(z)$ is the comoving source number density given the local source number density
$n_0$ and the cosmological evolution factor $\psi(z)$. 
Sources are distributed between redshift $z_{\rm min}$ and $z_{\rm max}$.
We define the burst rate per unit volume as $R_0$, and assume the local density
is effectively given by $n_0 = R_0\Delta T$. The diffuse cosmic background flux
in the present model is, therefore, described by ${\mathcal E}_\nu^{\rm fl}$, $R_0$, $\Delta T$,
and the evolution factor $\psi(z)$. The latter is
parametrized as $(1+z)^m$, the functional form often used in the literature.
The evolution factor for the transient neutrino sources is unknown, but
many of the proposed sources are related to SNe which approximately trace
the star formation rate (SFR). We therefore adopt the following parameterization, 
which approximately describes SFR~\citep{Yoshida:2014uka}, as our baseline model
\begin{eqnarray}
\psi(z) \propto \left\{ 
\begin{array}{ll}
(1 + z)^{3.4} & ( 0 \leq z \leq 1 ) \\
{\rm constant} & (1 \leq z \leq 4)
\end{array}
\right. .
\label{eq:sfd}
\end{eqnarray}
The diffuse cosmic background flux from the sources following an evolution factor
other than an SFR-like evolution
can be approximately estimated by scaling 
\begin{equation}
    E_\nu^2\Phi_\nu\approx E_\nu^2\Phi_\nu^{\rm SFR} \frac{\xi_z}{\xi_z^{\rm SFR}}
    \label{eq:evolution_scaling}
\end{equation}
where
\begin{equation}
    \xi_z=\frac{1}{H_0}\int\limits_{z_{\rm min}}^{z_{\rm max}}\frac{dz}{1+z}\left|\frac{dt}{dz}\right|\psi(z)
    \label{eq:effective_evolution_factor}
\end{equation}
is the effective evolution term and $\xi_z^{\rm SFR}\approx3$.

\begin{figure}[t!]
\plotone{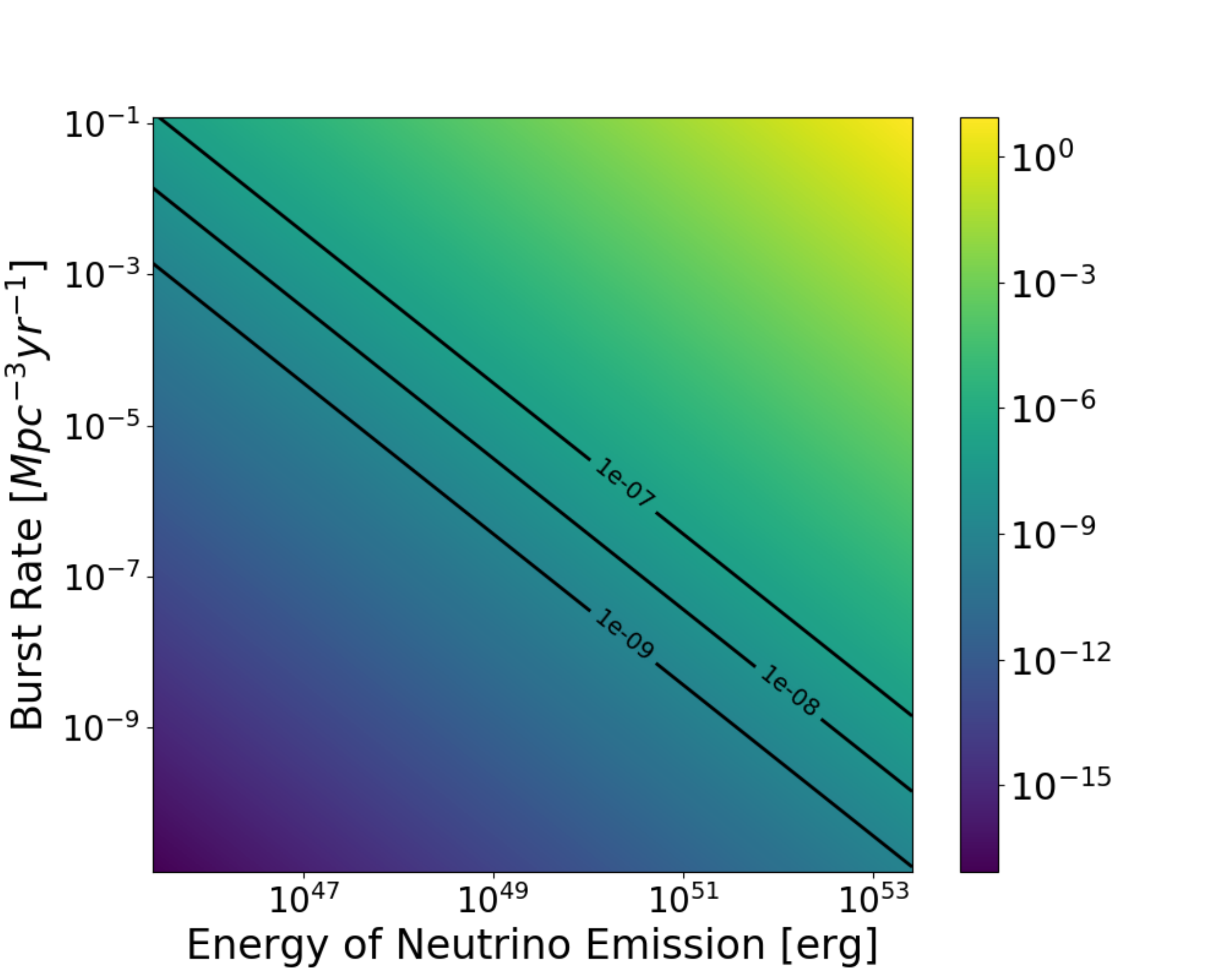}
\caption{The all-flavor-sum diffuse energy flux 
$\textstyle{E_\nu^2\Phi_\nu (3/\xi_z)}$ [${\rm GeV}\ {\rm cm}^{-2}\ {\rm s}^{-1}\ {\rm sr}^{-1}$] 
of the cosmic neutrino background radiation in
the $({\mathcal E}_\nu^{\rm fl}, R_0)$ plane, the total neutrino energy output
and the local burst density rate. The flare time duration $\Delta T= 30$~days
is assumed when calculating the local source number density.}
\label{fig:diffuse_flux}
\end{figure}
The resultant diffuse flux, $\Phi_\nu$, limits the range of the source parameters ${\mathcal E}_\nu^{\rm fl}$ and $R_0$ as the flux must be consistent with the IceCube measurements. Figure~\ref{fig:diffuse_flux} shows $\textstyle{E_\nu^2\Phi_\nu(3/\xi_z)}$ as a function of ${\mathcal E}_\nu^{\rm fl}$ and $R_0$. IceCube data suggests $E_\nu^2\Phi_\nu\lesssim 10^{-7}\ {\rm GeV}\ {\rm cm}^{-2}$~s$^{-1}$~sr$^{-1}$
(e.g. \citealt{Aartsen:2020aqd}).
Any parameter combination that yields $\textstyle{E_\nu^2\Phi_\nu}$ above this limit would overproduce diffuse flux. If
 $E_\nu^2\Phi_\nu\lesssim 10^{-9}\ {\rm GeV}\ {\rm cm}^{-2}$~s$^{-1}$~sr$^{-1}$,
the contribution to the TeV neutrino sky background would be negligible and so is not considered further.
Hence, we limit the source parameter space to meet
$10^{-9}\ {\rm GeV}\ {\rm cm}^{-2} {\rm s}^{-1} {\rm sr}^{-1} \leq E_\nu^2\Phi_\nu(3/\xi_z)\leq 10^{-7}\ {\rm GeV}\ {\rm cm}^{-2} {\rm s}^{-1} {\rm sr}^{-1}$.

\begin{figure*}[ht!]
\begin{center}
\includegraphics[width=0.32\textwidth]{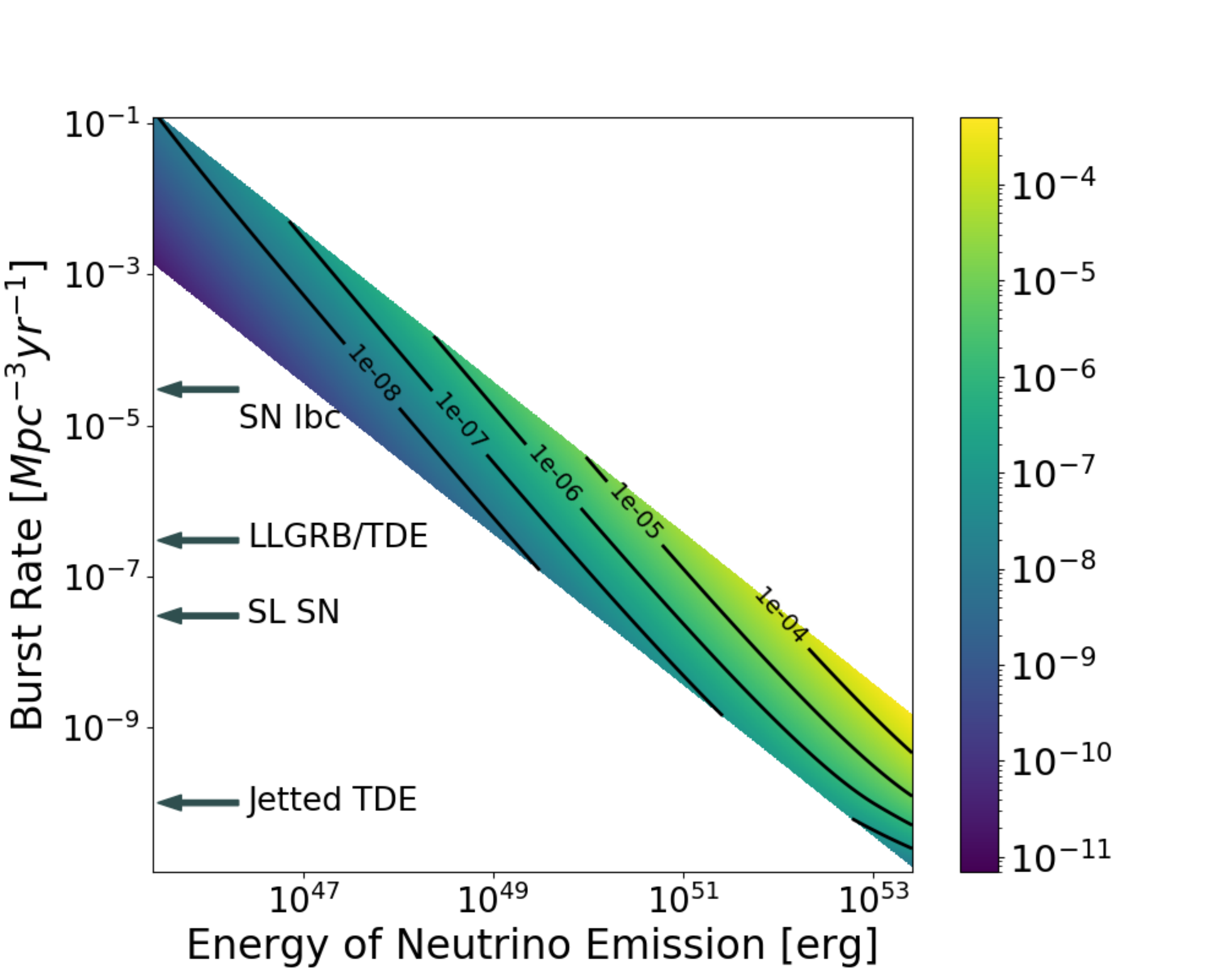}
\includegraphics[width=0.32\textwidth]{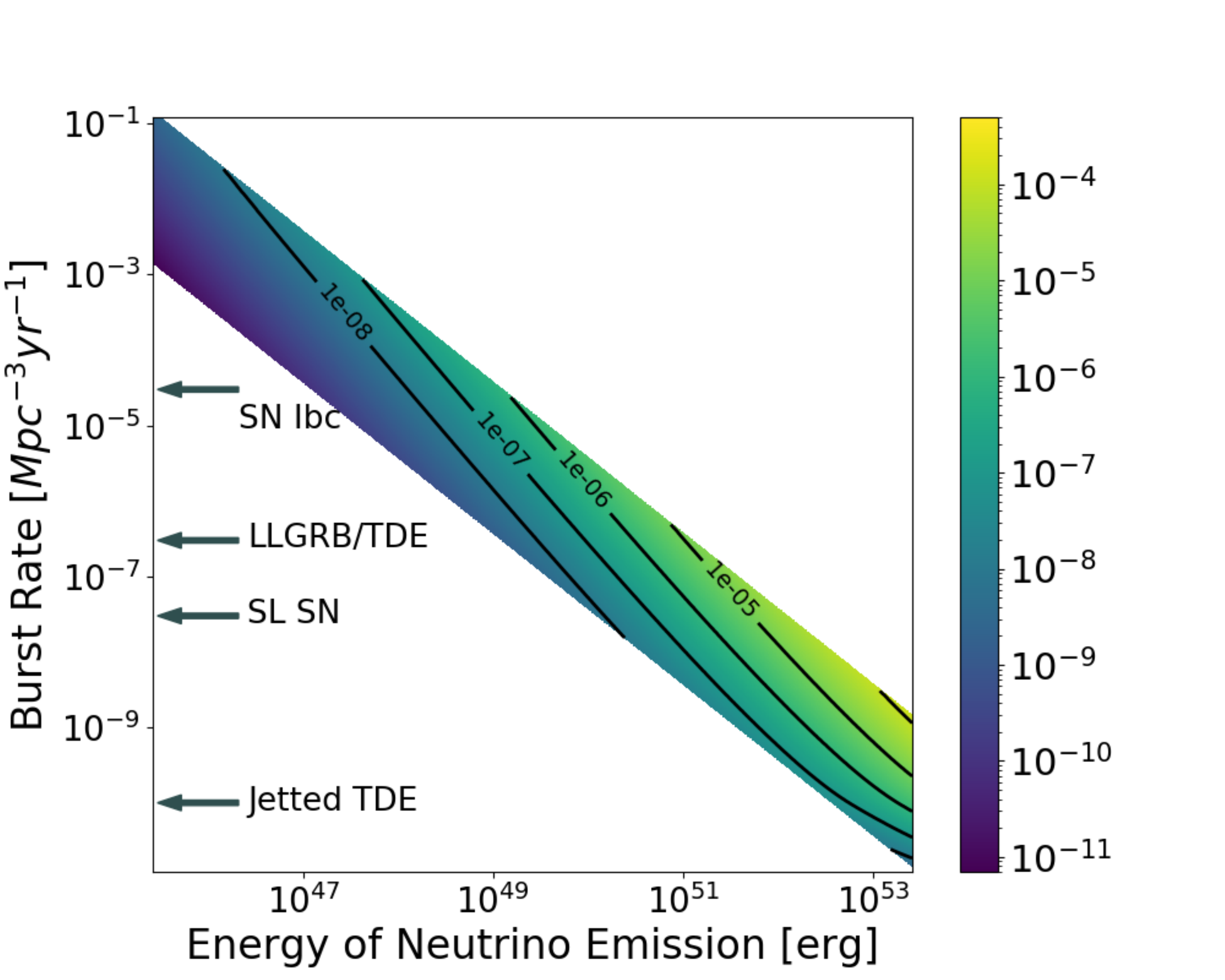}
\includegraphics[width=0.32\textwidth]{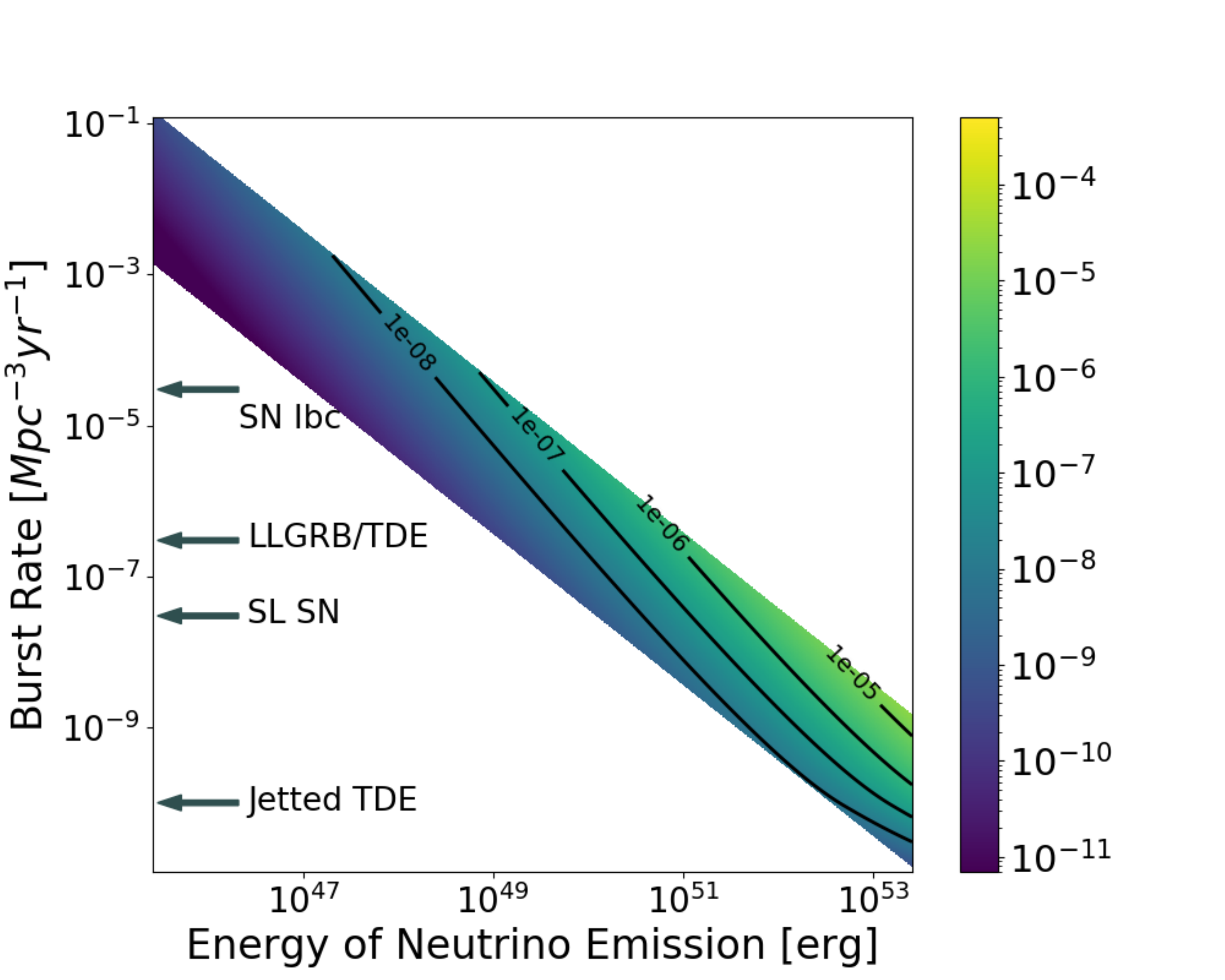}
\end{center}
\caption{(Left) Number of sources to yield neutrino multiplet, $N^{\rm M}_{\Delta \Omega}$,
in $\Delta \Omega= 1\ {\rm deg}^2$ of sky on the parameter space of $({\mathcal E}_\nu^{\rm fl}, R_0)$,
the output neutrino energy from a source
and the burst density rate. The expected ranges of the burst rate for several representative
transient source candidates are also shown for reference. 
(Middle) The effective p-value $P^{\rm eff}_{\rm m}$
to support statistically the hypothesis of multiplet detection from a transient source calculated by Eq.(\ref{eq:iceCube_multiplet_sensitivity}).
(Right) The effective number of multiplet sources given the p-value of the background-only hypothesis is
less than $10^{-6}$, which corresponds to an annual false alarm rate $\sim 0.25$ for the 2$\pi$ sky.}
\label{fig:number_of_multiplet_s}
\end{figure*}
The number of transient sources that could produce detectable neutrino multiplets for a given neutrino telescope is given by~\citep{Murase:2016gly} 
\begin{equation}
  N_{\Delta \Omega}^{\rm M}= \frac{\Delta\Omega}{4\pi}\int\limits_{z_{\rm min}}^{z_{\rm max}} dz d_z^2(1+z)\left|\frac{dt}{dz}\right| P_{\rm p}^{n\geq 2}[\mu^{\rm s}]n_0\psi(z),
  \label{eq:numbe_of_sources_multiplet}
\end{equation}
where $\Delta\Omega$ is the solid angle for a given direction of the neutrino multiplet and
$P_{\rm p}^{n\geq 2}$ is the Poisson probability of producing multiple neutrinos for the mean number of neutrinos $\mu^{\rm s}$ from a source at redshift $z$, given by
\begin{equation}
  \mu^{\rm s}(\Omega, z) = T_{\rm w}~\frac{1}{3}\int dE_\nu A_{\nu_\mu}(E_\nu, \Omega)\phi_\nu^{\rm fl}(E_\nu, z).
  \label{eq:event_ratefrom_PS}
\end{equation}
Note that the $1/3$ factor is applied for the conversion of the all-flavor-sum neutrino flux to that of per-flavor assuming the equal neutrino flavor ratio.
The muon neutrino detection effective area, $A_{\nu_\mu}$, determines the event rate.
We use an underground neutrino telescope model with a 1 km$^3$ detection volume 
~\citep{Gonzalez-Garcia:2009bev,Murase:2016gly} 
to estimate $A_{\nu_\mu}$ in our study.
Note that $P_{\rm P}^{\rm n=2}\sim \mu_{\rm s}^2/2 \propto d_z^{-4}$ when
$\mu^{\rm s}\ll 1$. Therefore, the integrand of $N^{\rm M}_{\Delta \Omega}\propto d_z^{-2}$, which indicates
that only nearby sources produce multiple events. $N^{\rm M}_{\Delta \Omega}$ is thus sensitive
to the minimal value of $d_z$, which is determined by $z_{\rm min}$. In our work,
it is defined as the average interval of source locations in the local universe, 
$\textstyle{r=(3/(4\pi n_0))^{1/3}}$.

In order to estimate the rate and the resultant sensitivity
for detecting the multiple neutrino events, we set the solid angle $\Delta\Omega$ to be comparable to the pointing resolution of
neutrino events. We assume $\Delta\Omega=1\ {\rm deg}^2$
hereafter, which is comparable to the angular resolution of track-like events
recorded by IceCube~\citep{IceCube:2013uto}.

The left panel of Figure~\ref{fig:number_of_multiplet_s} shows $N^{\rm M}_{\Delta \Omega}$ as a function of our source characterization parameters, ${\mathcal E}_\nu^{\rm fl}$ and $R_0$.
As $N^{\rm M}_{\Delta \Omega}$ represents the number of sources found within a solid angle of $\Delta\Omega$ 
during the multiplet search time frame $T_{\rm w}$, 
the expected number of sources to produce multiple neutrino events in the $2\pi$ sky detected in the observation time $T_{\rm obs}$ is given by
\begin{eqnarray}
    N_{\rm all}^{\rm M} &=& \left(\frac{2\pi}{\Delta\Omega}\right)\left(\frac{T_{\rm obs}}{T_{\rm w}}\right)
    N^{\rm M}_{\Delta \Omega} \nonumber\\
    &\simeq& 1.2 \left(\frac{T_{\rm obs}}{{\rm 5\ yr}}\right) \left(\frac{N^{\rm M}_{\Delta \Omega}}{10^{-6}}\right).
    \label{eq:rate_of_multiplet_sources}
\end{eqnarray}
Hence, the parameter space of 
$N^{\rm M}_{\Delta \Omega}\gtrsim 10^{-6}$
is accessible by a 1 km scale neutrino telescope~\footnote{i.e. the absence of neutrino multiples, $N_{\rm all}^{\rm M}\leq1$, can be used to constrain the population of neutrino source candidates~\citep{Murase:2016gly,Ackermann:2019ows}.}.

Note that ${\mathcal E}_\nu^{\rm fl}=L_\nu\Delta T$ represents the total energy of neutrinos if $T_{\rm w}>(1+z)\Delta T$. In general, $L_\nu T_{\rm w}$ can be smaller than ${\mathcal E}_\nu^{\rm fl}$ if $T_{\rm w}<(1+z)\Delta T$. 

\subsection{Estimates  of the sensitivity of the multiplet detection against the atmospheric background} \label{subsec:multiplet_likelihood analysis}


The number of sources to produce multiplet events $N^{\rm M}_{\Delta \Omega}$ given by
Eq.~(\ref{eq:numbe_of_sources_multiplet}) is equivalent
to the multiplet signal rate from the viewpoint of a neutrino detector.
However, searches for neutrino multiple events from these sources are contaminated
by the atmospheric and the unresolved diffuse cosmic neutrino backgrounds. The expected atmospheric background for $T_{\rm w}=30$~days 
and $\Delta\Omega=1$~deg$^2$
reaches $\sim 0.5$ events in the solid angle average for a km$^3$ volume neutrino detector,
which would smear out the neutrino signals from the source. 
Because the atmospheric neutrino spectrum
follows $\sim E_\nu^{-3.7}$ which is much softer than 
$\phi_\nu^{\rm fl}\propto E_\nu^{-\alpha_\nu}$, 
taking into account the energy of each of the multiple neutrino events 
can adequately remove the background contamination.
For this purpose, we construct the extended Poisson likelihood functions
for the signal hypothesis $\mathcal{L}^{\sf sig}$ to describe
the case when the detected multiple events originate in transient sources,
and $\mathcal{L}^{\sf BG}$ for the background hypothesis as
\begin{eqnarray}
  \mathcal{L}^{\sf sig} &=&
  (1-e^{-N^{\rm M}_{\Delta \Omega}})e^{-(\mu_{\rm atm}+\mu_{\rm dif})}\prod_{i=i}^N P_{\rm s}^{\rm E}(E_\nu^i). \nonumber\\
  \mathcal{L}^{\sf BG} &=&
  e^{-N^{\rm M}_{\Delta \Omega}}\left(1-e^{-(\mu_{\rm atm}+\mu_{\rm dif})}- \right. \nonumber\\
 && \left. (\mu_{\rm atm}+\mu_{\rm dif})e^{-(\mu_{\rm atm}+\mu_{\rm dif})}\right) \nonumber\\
 && \prod_{i=1}^N \left[ \frac{\mu_{\rm atm}}{\mu_{\rm atm}+\mu_{\rm dif}}P_{\rm atm}^{\rm E}(E_\nu^i) +
   \frac{\mu_{\rm dif}}{\mu_{\rm atm}+\mu_{\rm dif}}P_{\rm dif}^{\rm E}(E_\nu^i)\right]. \nonumber\\
    \label{eq:extended_likelihood_final}
\end{eqnarray}
Here $\mu_{\rm atm}\simeq 0.5$, and $\mu_{\rm dif}\lesssim 10^{-3}$ are the expected mean number events
from the atmospheric and the unresolved diffuse neutrino backgrounds, respectively.
$P_{\rm s}^{\rm E}, P_{\rm atm}^{\rm E}$, and $P_{\rm dif}^{\rm E}$ are 
the probability density functions (pdf) of the
energies of neutrinos from the transient sources, the atmospheric background, and the diffuse flux.
These are obtained by multiplying $A_{\nu_\mu}$ with each of the neutrino fluxes.
In the limit $N_{\rm PS}^{\rm M}\ll 1$ and $\mu_{\rm atm}+\mu_{\rm dif} \ll 1$ where we
consider only the doublet case,
this likelihood can be described by a more intuitive formulas
\begin{eqnarray}
  \mathcal{L}^{\sf sig} &\simeq& N^{\rm M}_{\Delta \Omega}\prod_{i=i}^2 P_{\rm s}^{\rm E}(E_\nu^i) \nonumber\\
  \mathcal{L}^{\sf BG}  &\simeq& (1-N^{\rm M}_{\Delta \Omega})\frac{1}{2}
  \prod_{i=1}^2 \left[ \mu_{\rm atm}P_{\rm atm}^{\rm E}(E_\nu^i) +\mu_{\rm dif}P_{\rm dif}^{\rm E}(E_\nu^i)\right]. \nonumber\\
    \label{eq:extended_likelihood_limit}
\end{eqnarray}

\begin{figure}[t!]
\plotone{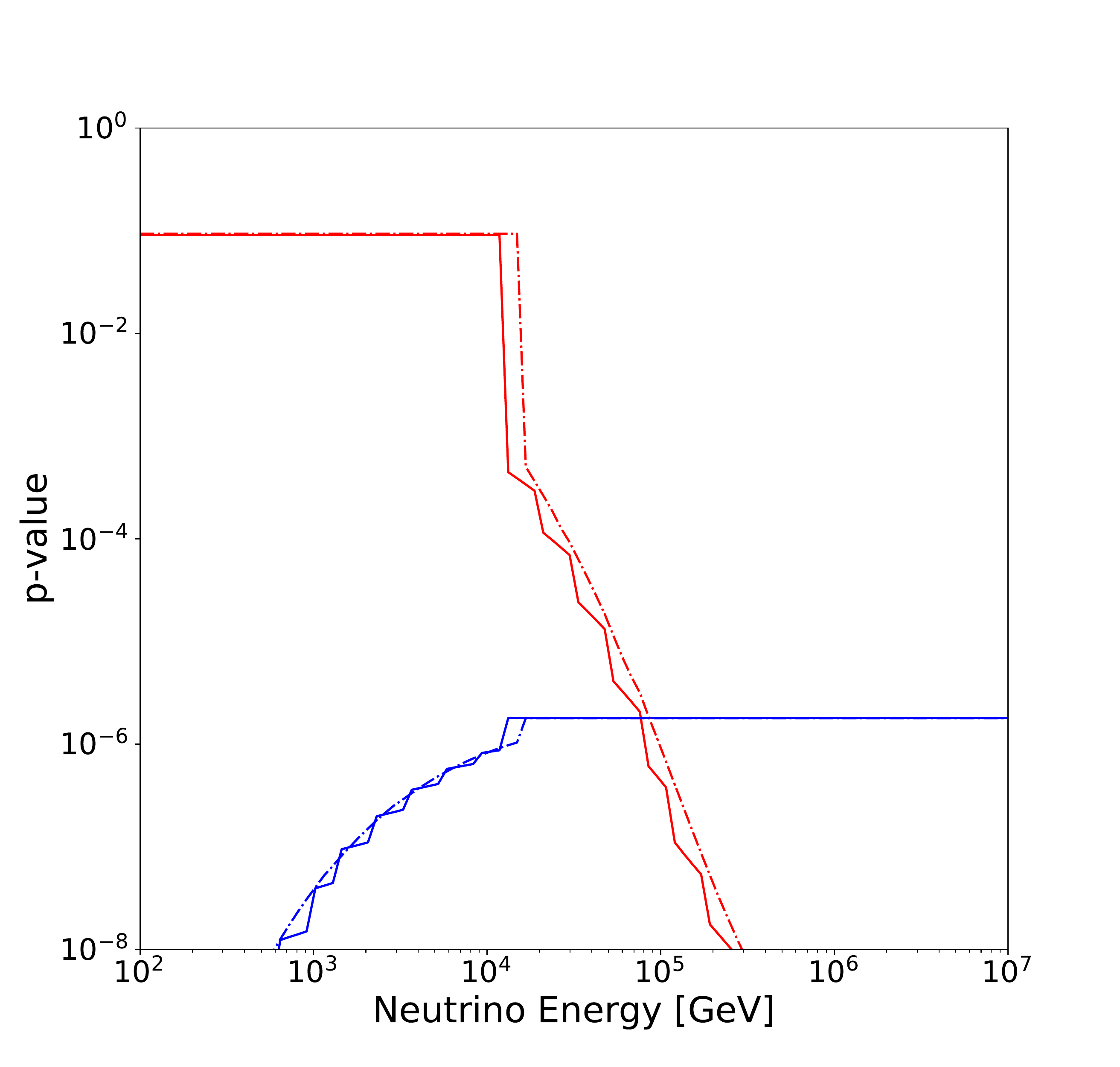}
\caption{P-values for the background-only hypotheses (red) and the transient signal hypothesis
with $N^{\rm M}_{\Delta \Omega}=3\times 10^{-6}$ (blue) as a function of the doublet neutrino energy
detected from a $\Delta\Omega= 1 {\rm deg}^2$ patch of sky for the time interval
$T_{\rm w}=30\ {\rm days}$.
The dashed curves show the case when the error of neutrino energy estimation is assumed to be
$\sigma_{\rm \log E} = 0.2$. See text for details.}
\label{fig:p-values-doub}
\end{figure}

The test statistic for the background-only hypothesis is constructed with the log likelihood ratio
\begin{equation}
  \Lambda = 2\ln\frac{\mathcal{L}(\widehat{N^{\rm M}_{\Delta \Omega}}({\mathcal E}_\nu^{\rm fl},R_0))|_{{\sf BG, sig}}}{\mathcal{L}^{\sf BG}(N^{\rm M}_{\Delta \Omega}=0)},
  \label{eq:TS_icecube_bgOnly}
\end{equation}
where the hat notation represents the value that maximizes the likelihood.
The test statistic for the signal hypothesis with $({\mathcal E}_\nu^{\rm fl}, R_0)$
is 
\begin{equation}
  \Lambda = 2\ln\frac{\mathcal{L}(\widehat{N^{\rm M}_{\Delta \Omega}}({\mathcal E}_\nu^{\rm fl},R_0))|_{{\sf BG, sig}}}{\mathcal{L}^{{\sf sig}}(N^{\rm M}_{\Delta \Omega}({\mathcal E}_\nu^{\rm fl}, R_0))}.
  \label{eq:TS_icecube}
\end{equation}
Figure~\ref{fig:p-values-doub} shows the calculated p-values as a function of neutrino energy
in a doublet, $E_\nu = E_\nu^1=E_\nu^2$. We set $N^{\rm M}_{\Delta \Omega}$ = $3\times 10^{-6}$
for illustrative purpose.
The p-values that support the signal hypothesis
reach a plateau 
$\textstyle{e^{-(\mu_{\rm atm}+\mu_{\rm dif})}N_{\Delta \Omega}^{\rm M}\simeq 1.8\times 10^{-6}}$ 
when the doublet neutrino energy
gets higher, as expected. The $2\pi$ sky converted annual false alarm rate (FAR) is $\sim 0.25~ \mathrm{yr}^{-1}$ 
when the doublet energy is higher than $\varepsilon_0=100$~TeV.
Note that the intensity of the prompt atmospheric neutrino component~\citep{Bhattacharya:2015jpa}
produced from the decay of heavy charmed hadrons was found to be subdominant 
compared to that of the astrophysical neutrinos~\citep{IceCube:2021uhz},
and the statistical test discussed here is robust against the uncertainties originating
in the heavy hadron physics.

The sensitivity for a given $N_{\rm PS}^{\rm M}({\mathcal E}_\nu^{\rm fl}, R_0)$ is evaluated
by convolution of p-values for the multiplet source hypothesis, 
given by the test statistic Eq.~(\ref{eq:TS_icecube}),
with the probability of detecting $E_\nu$ energy neutrinos
\begin{equation}
  P^{\rm eff}_{\rm m} = \int dE_\nu^1 P_{\rm s}^{\rm E}(E_\nu^1)\int dE_\nu^2
  P_{\rm s}^{\rm E}(E_\nu^2)p(E_\nu^1, E_\nu^2),
  \label{eq:iceCube_multiplet_sensitivity}
\end{equation}
where $p(E_\nu^1, E_\nu^2)$ represents the p-value for doublet $(E_\nu^1,E_\nu^2)$.
This indicates the {\it effective} rate (or p-value) of finding a source 
that produces multiple neutrino events, 
considering the atmospheric background contamination.
The middle panel of Figure~\ref{fig:number_of_multiplet_s}
shows the effective p-values, $P^{\rm eff}_{\rm m}$. For a given neutrino source parameter set
$({\mathcal E}_\nu^{\rm fl}, R_0)$, this shows how often we can see multiple
neutrinos that are inconsistent with the atmospheric neutrino background
for $\Delta\Omega=1$~deg$^2$ and $T_{\rm w}=$~30 days. The domain 
of $P^{\rm eff}_{\rm m}\gtrsim 10^{-6}$ is reachable by a five year observation
with $2\pi$ sky coverage.

A null detection of any multiple neutrino events by a 1 km$^3$ neutrino telescope
constrains the neutrino source parameters $({\mathcal E}_\nu^{\rm fl}, R_0)$.
Hence, criteria for rejecting the neutrino background-only hypothesis globally 
across the $2\pi$ sky must be adequately introduced. 
As an example, in the right panel of Figure~\ref{fig:number_of_multiplet_s}, 
$N^{\rm M}_{\Delta \Omega}$ requires the criterion that 
with the local p-value of a neutrino multiplet for the background-only hypothesis must be
less than $10^{-6}$. This corresponds to an annual global FAR of $\sim 0.25$
in the $2\pi$ sky. 
The parameter space where $N^{\rm M}_{\Delta \Omega}\gtrsim 10^{-6}$
can be constrained by a few years of observations when the rate of multiplet detection
under this condition is consistent with the global FAR.

\section{Optical followup observations} \label{sec:followup}

\subsection{Statistical strategy to reject unrelated supernovae} \label{subsec:statstic_followup}


Neutrino sources that yield multiple neutrino events are likely to be optical transient objects. Searching for optical/NIR counterparts in followup observations to identify the neutrino emitter can be contaminated by the detection of more dominant SNe. We anticipate finding $\gtrsim 100$ SNe in a field of view of $\Delta\Omega=1\ {\rm deg}^2$ during a time window of $T_{\rm w}=30$~days.
Their redshift distribution is, however, quite different from that expected for the neutrino multiplet sources. As we have already shown in Figure~\ref{fig:NSources4intro}, most of the sources that produce multiple neutrino events are confined within the local universe.
The resultant difference in the probability distribution of redshift between the neutrino multiplet sources and the unrelated SNe
allows for a statistical test to indicate
which hypothesis is more consistent with the observational data. 

Among the candidate optical transient counterparts, the object with the minimum redshift, $\scalebox{1.2}{\it z}_{\scalebox{0.6}{ min}}^{\scalebox{0.6}{ trans}}$, is the most likely source of the neutrino multiplet event. The pdf of $\scalebox{1.2}{\it z}_{\scalebox{0.6}{ min}}^{\scalebox{0.6}{ trans}}$,  
which is defined by our signal hypothesis,
is obtained by normalizing $N^{\rm M}_{\Delta \Omega}$ from Eq.~(\ref{eq:numbe_of_sources_multiplet}):
\begin{equation}
  \rho_{z_{\rm min}}^{\rm M}({\mathcal E}_\nu^{\rm fl}, R_0, \scalebox{1.2}{\it z}_{\scalebox{0.6}{ min}}^{\scalebox{0.6}{ trans}}) = \frac{1}{N^{\rm M}_{\Delta \Omega}}\frac{dN^{\rm M}_{\Delta \Omega}}{dz}(z=\scalebox{1.2}{\it z}_{\scalebox{0.6}{ min}}^{\scalebox{0.6}{ trans}})
  \label{eq:mS_PDF}
\end{equation}
and the likelihood is constructed by 
$\mathcal{L}_{\rm S}^{\rm M} (\scalebox{1.2}{\it z}_{\scalebox{0.6}{ min}}^{\scalebox{0.6}{ trans}}) = \rho_{z_{\rm min}}^{\rm M}({\mathcal E}_\nu^{\rm fl}, R_0, z_{\rm min}^{\rm tran})$

In the background hypothesis case, the closest counterpart object belongs
to the population of SNe not associated with the neutrino detection.
The number of unrelated SNe within redshift $z$
in a field of view of $\Delta \Omega$ is given by
\begin{equation}
  N_{\rm SN}(z)=\frac{\Delta\Omega}{4\pi} n^{\rm SN}_0 \int\limits_0^z dz
  d_z^2(1+z)\left|\frac{dt}{dz}\right|\psi_{\rm SN}(z).
  \label{eq:sN_number}
\end{equation}
Here, $\psi_{\rm SN}(z)$ is the cosmological evolution term, which is assumed to follow an SFR-like distribution, represented by Eq.~(\ref{eq:sfd}),
and $n^{\rm SN}_0$ is the average number density of SNe in the present epoch, which is effectively obtained from the SNe rate, $R^{\rm SN}_0$, as
\begin{equation}
  n^{\rm SN}_0 = R^{\rm SN}_0 T_{\rm w}.
  \label{eq:SNrateToDensity}
\end{equation}
$R^{\rm SN}_0=1.3\times 10^{-4}$ Mpc$^{-3}$~yr$^{-1}$ is assumed as the nominal value 
in our work (\citealt{madau14} and references therein).
The pdf of its redshift is given by
\begin{equation}
  \rho_z^{\rm SN}(z) = \frac{1}{N_{\rm SN}}\frac{dN_{\rm SN}}{dz}.
  \label{eq:SN_PDF}
\end{equation}
The pdf of $\scalebox{1.2}{\it z}_{\scalebox{0.6}{ min}}^{\scalebox{0.6}{ trans}}$ in the background hypothesis is
given by
\begin{eqnarray}
  \rho_{z_{\rm min}}^{\rm SN} &=& \left(1-\int_0^{\scalebox{1}{\it z}_{\scalebox{0.5}{ min}}^{\scalebox{0.5}{ trans}}} dz \rho_z^{\rm SN}(z) \right)^{N_{\rm SN}-1}N_{\rm SN}\rho_z^{\rm SN}(\scalebox{1.2}{\it z}_{\scalebox{0.6}{ min}}^{\scalebox{0.6}{ trans}})\nonumber \\
  &\simeq& \exp\left({-\int_0^{\scalebox{1}{\it z}_{\scalebox{0.5}{ min}}^{\scalebox{0.5}{ trans}}} dz \frac{dN_{\rm SN}}{dz}}\right) \frac{dN_{\rm SN}}{dz}(\scalebox{1.2}{\it z}_{\scalebox{0.6}{ min}}^{\scalebox{0.6}{ trans}}).
  \label{eq:SN_pdf}
\end{eqnarray}
The likelihood for the background hypothesis is constructed by 
$\mathcal{L}_{\rm BG}^{\rm SN} (\scalebox{1.2}{\it z}_{\scalebox{0.6}{ min}}^{\scalebox{0.6}{ trans}}) = \rho_{z_{\rm min}}^{\rm SN}(R^{\rm SN}_0, \scalebox{1.2}{\it z}_{\scalebox{0.6}{ min}}^{\scalebox{0.6}{ trans}})$. 

\begin{figure}[t!]
\plotone{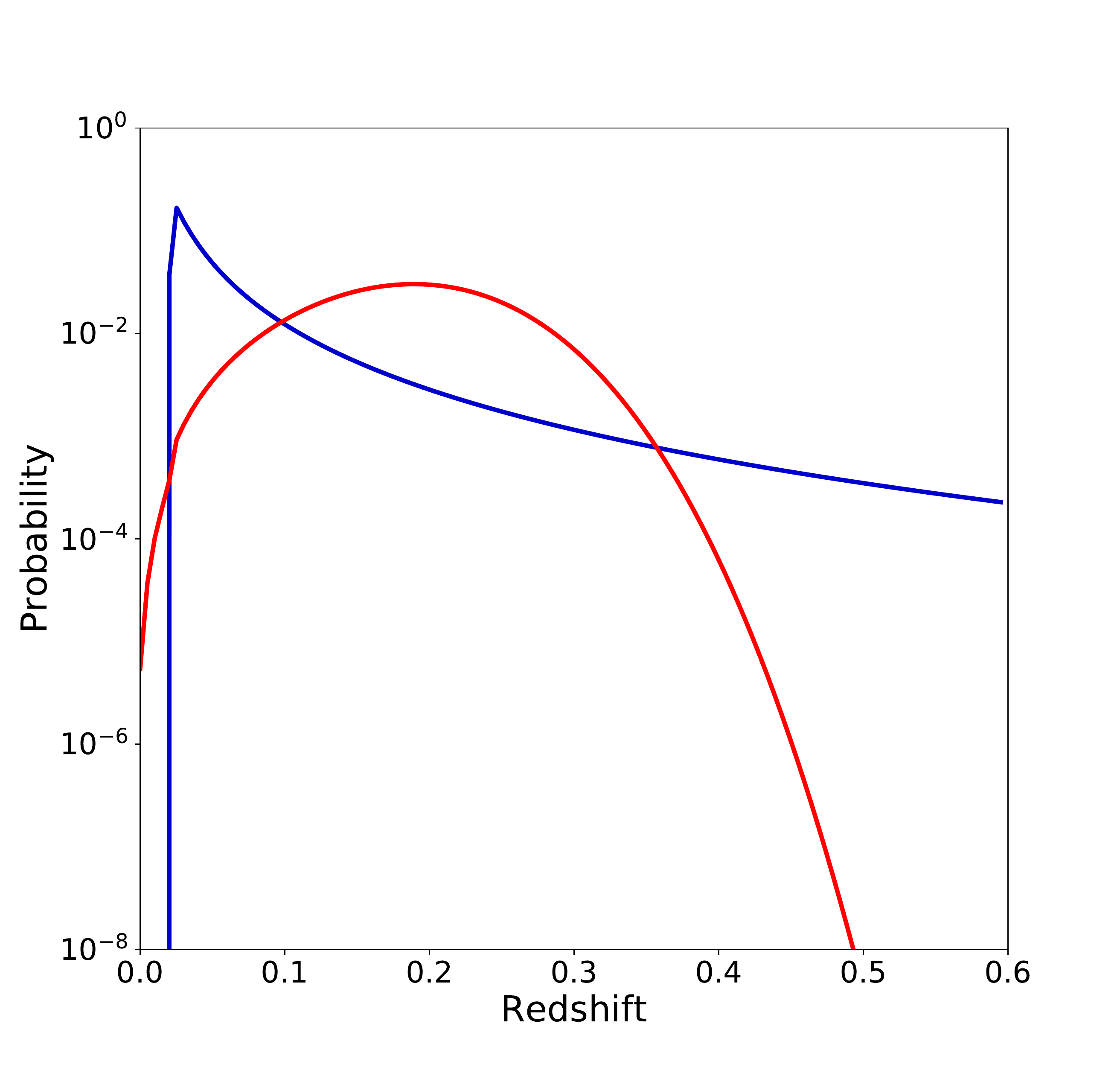}
\caption{Probability distribution of $z_{\min}^{\rm trans}$ as a function of redshift with bin size of
$\Delta z = 0.005$. The bin size is chosen for illustrative purposes. 
The blue curve represents the case of the signal hypothesis, and the red curve shows the case
of the coincident background hypothesis. ${\mathcal E}_\nu^{\rm fl}=1\times 10^{49}$~erg and 
$R_0=3\times 10^{-6}$~Mpc$^{-3}$~yr$^{-1}$ are assumed for the multiplet source.
The average SNe rate $R^{\rm SN}_0=1.3\times 10^{-4}$~Mpc$^{-3}$~yr$^{-1}$
in the calculation of $\rho_{z_{\rm min}}^{\rm SN}$ while
$R^{\rm SN, local}_0=7.0\times 10^{-5}$~Mpc$^{-3}$~yr$^{-1}$ 
is assumed in the local universe within a 100 Mpc radius. See the main text for details.}
\label{fig:redshift_pdf}
\end{figure}
Figure~\ref{fig:redshift_pdf} shows the probability distribution of $\scalebox{1.2}{\it z}_{\scalebox{0.6}{ min}}^{\scalebox{0.6}{ trans}}$
at a given redshift for the signal and the background hypothesis, respectively. 
Note that the lower bound of the allowed redshift for the signal hypothesis
is determined by the average source distance interval $\textstyle{r=(3/(4\pi n_0))^{1/3}}$.
The substantial difference between the two distributions
provides a statistical power to calculate the p-values to reject the unrelated SN hypothesis
for a given $\scalebox{1.2}{\it z}_{\scalebox{0.6}{ min}}^{\scalebox{0.6}{ trans}}$.
The test statistic to reject the background hypothesis is constructed by
\begin{equation}
    \Lambda = \ln \frac{\mathcal{L}_{\rm S}^{\rm M}}{\mathcal{L}_{\rm BG}^{\rm SN}}.
\end{equation}
Table~\ref{tab:p-value_tbl} lists the p-values for the background hypothesis.
If we find the closest counterpart at redshift of 0.05, the statistical significance
against the incorrect coincident SN detection hypothesis is $\sim 2.4\sigma$.
Any distant counterpart with $z\gtrsim 0.15$ would exhibit a $\sim 1\sigma$
significance at most, and its association with the neutrino multiplet 
cannot be claimed.

\begin{table}[tb]
    \centering
    \caption{p-value for the background hypothesis when the closest optical transient counterpart 
    is at redshift $z_{\rm min}^{\rm tran}$. 
    The two cases of the local B-band luminosity are shown for reference.}
    \begin{tabular}{lcc}
    \hline\hline
       $\scalebox{1.2}{\it z}_{\scalebox{0.6}{ min}}^{\scalebox{0.6}{ trans}}$ & p-value  & p-value  \\ 
        & $\rho_{\rm B}=1\times10^8 L_\odot$ & $\rho_{\rm B}=1\times10^9 L_\odot $ \\ 
       \hline
       0.03  & $7.5\times 10^{-4}$ & $9.8\times 10^{-4}$ \\
       0.04  & $3.4\times 10^{-3}$ & $3.0\times 10^{-3}$\\
       0.05  & $9.3\times 10^{-3}$ & $9.0\times 10^{-3}$\\
       0.1  &  $7.7\times 10^{-2}$ & $7.7\times 10^{-2}$\\
       0.15 &  $3\times 10^{-1}$ & $3\times 10^{-1}$\\
       \hline
    \end{tabular}
    \label{tab:p-value_tbl}
\end{table}
The SNe rate may not be exactly a volume-averaged value. 
The inhomogeneous distribution of galaxies in our local universe 
impacts the expected value of the nearby SNe rate. For example, the local SFR suggests that the CC SNe rate within 10 Mpc is larger than the volume-averaged value, which enhances the detectability of high-energy neutrinos from nearby SNe \citep{Kheirandish:2022eox}. 
The local anisotropic and inhomogeneous structure appears on a distance scale of $\lesssim$ 100 Mpc, governed by the local clusters of galaxies. 
In order to evaluate the robustness of the present statistical approach,
we build an empirical local universe model to calculate the local SNe rate
within 100 Mpc of our Galaxy.

The local SFR $R^{\rm SFR}_0$ is associated with the average B-band luminosity $\bar{L}_{\rm B}$ \citep{kennicutt98}. 
We assume $\textstyle{R^{\rm SN}_0 \propto R^{\rm SFR}_0\propto \bar{L}_{\rm B}}$.
The ratio of $R^{\rm SFR}_0$ and $\bar{L}_{\rm B}$ is approximately constant for various galaxy types \citep{james08}:
\begin{equation}
  \frac{R^{\rm SFR}_0}{\bar{L}_{\rm B}} \simeq 10^{-10\pm 0.5}\frac{M_\odot\ {\rm yr}^{-1}}{L_\odot}.
  \label{eq:SFRperB}
\end{equation}
$R^{\rm SN}_0$ in the local universe, $R^{\rm SN, local}_0$, 
is related to  $R^{\rm SFR}_0$ via
\begin{equation}
  R^{\rm SN, local}_0 = \frac{1}{M_{\rm SN}}\frac{R^{\rm SFR}_0}{\bar{L}_{\rm B}}\rho_{\rm B}.
  \label{eq:localSNRate}
\end{equation}
Here $\rho_{\rm B}$ is the B-band luminosity density of the local universe, 
and $1/M_{\rm SN}$ corresponds to the probability of SNe per mass, 
which is given by
\begin{eqnarray}
  \frac{1}{M_{\rm SN}}&=&\frac{\int\limits_{M_{{\rm min}}^{\rm SN}}dM_s \frac{dN_s}{dM_s}}{\int\limits_{M_{{\rm min}}}dM_s M_s \frac{dN_s}{dM_s}}\nonumber\\
  &\sim& 7\times 10^{-3}M_\odot^{-1},
\end{eqnarray}
where $M_{{\rm min}}$ and $M_{{\rm min}}^{\rm SN}$ are the minimum masses for the stellar population and for SNe, respectively. We assume $M_{\rm min} = 0.1 M_{\odot}$ and 
$M_{\rm min}^{\rm SN} = 8 M_{\odot}$
by following \citet{madau14}.
For the initial mass function, $dN_s/dM_s \propto M_s^{-2.35}$ is assumed.

\begin{figure}[tb!]
\plotone{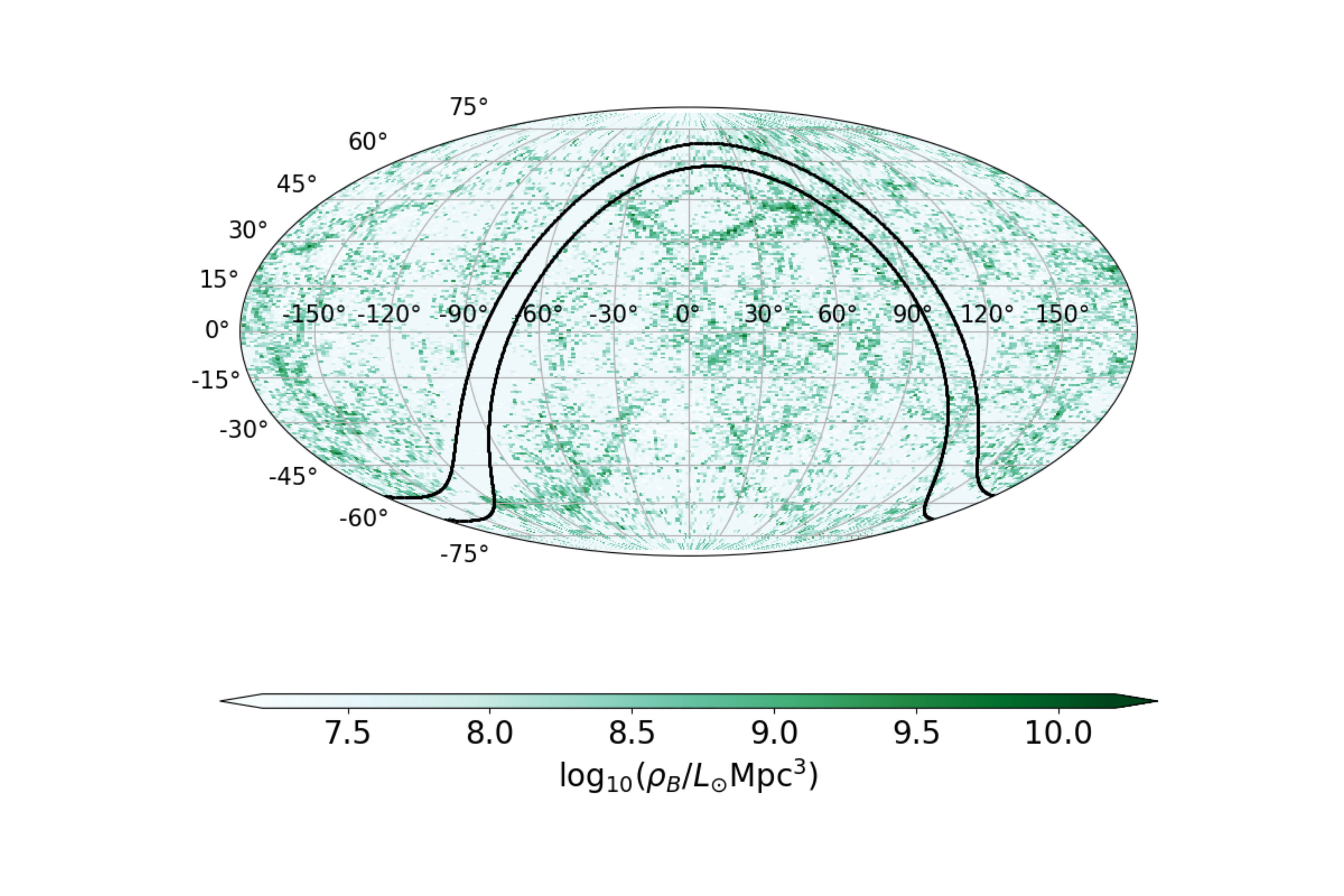}
\caption{Skymap of the B-Band luminosity density in equatorial coordinates. The band defined
by the black lines is removed from the density profile calculations due to the contamination
associated with the Galactic plane. The region where no galaxy is registered
within 100 Mpc is set to be $2.4\times 10^7 L_\odot$ Mpc$^{-3}$ for
the conservative estimate determined by the catalogue completeness.
\label{fig:B-band-skymap}}
\end{figure}
We consider the B-band luminosity density $\rho_{\rm B}$ for a given patch of the local universe 
as a measure to quantify a departure from the cosmologically averaged star formation activity.
It is calculated by 
\begin{equation}
  \rho_{\rm B}= \frac{1}{V_{\rm local}}\int dn_{\rm gal}L_{\rm B}.
  \label{eq:bBandLuminosityDensity}
\end{equation}
The galaxy number distribution $dn_{\rm gal}/dL_{\rm B}$ can be estimated
by the actual observations of galaxies.
By using the GLADE galaxy catalog \citep{dalya18}, 
we estimated $\rho_{\rm B}$ within 100 Mpc for various directions with $\Delta\Omega = 1$ deg$^2$
by counting galaxies brighter than -18 mag which results in $\sim 60$ \%  catalog completeness
in terms of the integrated B-band luminosity. Figure~\ref{fig:B-band-skymap} 
shows the estimated $\rho_{\rm B}$ skymap in equatorial coordinates.
The mean number was found to be
$\rho_{\rm B} \simeq 10^8 L_\odot$ Mpc$^{-3}$ while
$\rho_{\rm B} \lesssim 10^9 L_\odot$ Mpc$^{-3}$ in 98\% of the sky patches.
We take the latter value as the upper bound for conservative estimates
of the p-values for the background SNe hypothesis.

When $10^8 L_\odot\ {\rm Mpc}^{-3}\lesssim \rho_{\rm B} \lesssim 10^9 L_\odot$ Mpc$^{-3}$,
$R^{\rm SN, local}_0$ calculated by Eq.~(\ref{eq:localSNRate}) ranges from $7\times 10^{-5}$~Mpc$^{-3}$~yr$^{-1}$
to $7\times 10^{-4}$~Mpc$^{-3}$~yr$^{-1}$. Table~\ref{tab:p-value_tbl} shows
the p-value for these two cases for the local universe, for the average and upper bounds of
the galaxy density.

\subsection{Identification of a transient associated with neutrino multiplet} \label{subsec:observation_followup}



\begin{figure}[tb!]
\plotone{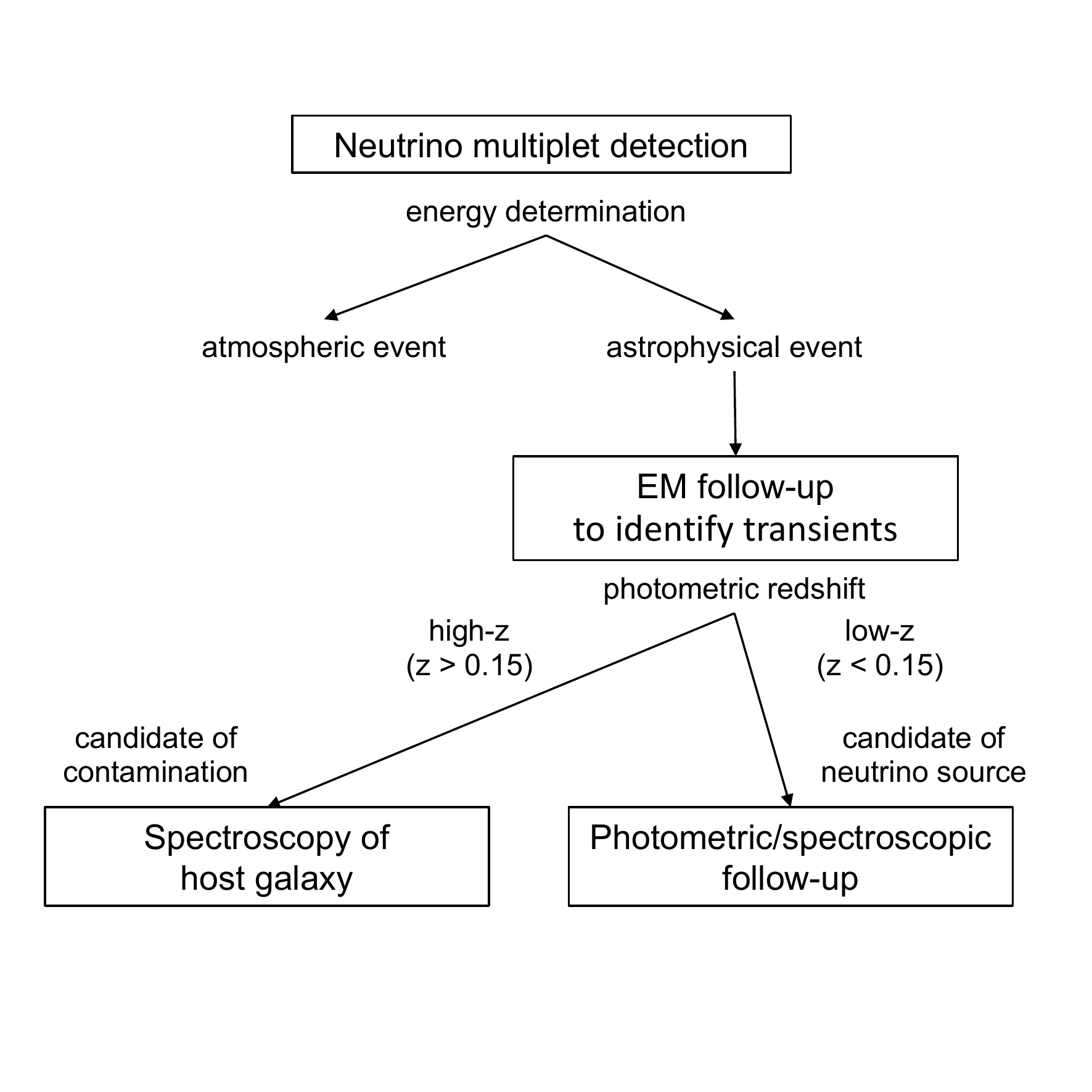}
\caption{The flow chart of the procedures to find a neutrino source candidate
following a neutrino multiplet detection.
\label{fig:flowchart}}
\end{figure}

As described in Section \ref{subsec:statstic_followup}, 
the multiplet source is most likely located at
a relatively small redshift ($z < 0.15$ with an 88\% probability).
Given that most astrophysical neutrino multiplets will therefore originate in objects with small redshifts, we can construct a follow-up search strategy  for identifying the optical counterpart.
Figure~\ref{fig:flowchart} shows
a flow chart to illustrate our proposed procedure to find a source candidate.
The first step in this procedure for the followup observations 
is to see if the optical transient counterparts are close enough. We propose $z=0.15$
as the threshold for the first level selection.

\begin{figure}[tb!]
\plotone{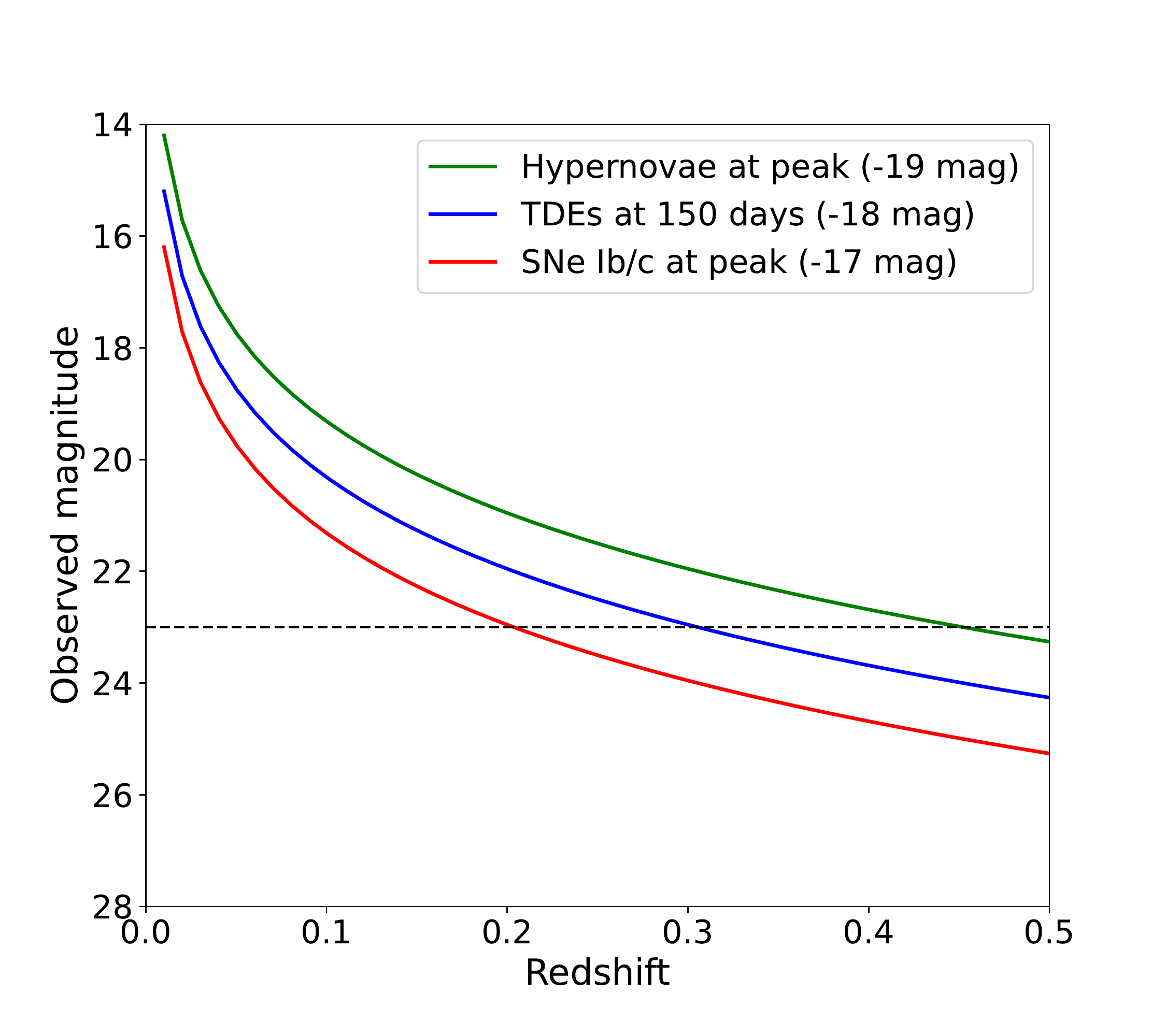}
\caption{Observed magnitude as a function of redshift for target objects 
after expected time delays
(green: hypernovae at peak; red: TDEs at 100 days after the event; blue: SNe Ibc at peak).
\label{fig:mag_redshift}}
\end{figure}

To detect the optical emission from transients at $z \lesssim 0.15$, 
the required sensitivity for optical follow-up observations is about 23~mag.
Figure \ref{fig:mag_redshift} shows the typical peak optical magnitude of different transients 
as a function of redshift.
The sensitivity of 23~mag is sufficient to detect hypernovae
and broad-lined, energetic Type Ic SNe at $z = 0.15$.
To perform such an optical survey for 1 deg$^2$ area, 
wide-field optical telescopes with a diameter of $>$ 4~m, such as DECam on the 4 m Blonco telescope \citep{flaugher15}, HSC on the 8.2~m Subaru telescope \citep{miyazaki18}, and the 8.4~m Rubin observatory telescope \citep{ivezic19}, are required.

Such a deep survey can identify not only the true multiplet source but also unrelated transient objects (i.e., contaminants).
In the following sections, we first estimate the number of contamination sources that may be found with such a survey,
and then discuss the strategy to identify the neutrino multiplet source.

\subsubsection{Expected number of contaminants}
\label{subsubsec:contaminants}
We perform survey simulations 
to estimate the number of contaminants in an observation with
23~mag sensitivity. 
Most optical transients are common SNe, 
namely Type Ia SNe (thermonuclear SNe) and Type II SNe, and Ibc SNe (CC SNe).
These SNe are located by successive imaging observations. Our strategy 
is to survey the neutrino direction sky patch of $\Delta\Omega=1$~deg$^2$
three times with a time interval comparable to the typical timescales of light curves of SNe. 

The light curves of normal SNe are generated
with the {\tt sncosmo} package \citep{barbary16} 
by using templates of the available spectral energy distributions and there time evolution. 
For Type Ia SNe, the SALT2 model \citep{guy07} is used as a template.
The parameters of the SALT2 model, stretch and color parameters, are randomly selected
according to the measured distribution \citep{scolnic16}.
The peak luminosity of Type Ia SNe is assigned from the stretch and color parameters.
For Type II and Ibc SNe, the set of the spectral templates in {\tt sncosmo} is used.
These spectral energy distribution templates are based on each type of observed SN, 
and therefore, we randomly select the template SNe to generate the light curves.
The distribution of the peak luminosity of Type II and Ibc SNe is assumed to be 
Gaussian with an average absolute magnitude of $-16.80$~mag and $-17.50$~mag 
and a dispersion $0.97$~mag and $1.10$~mag for Type II and Ibc SNe, respectively \citep{richardson14}.
Although the true luminosity distribution is not a Gaussian distribution \citep{perley20}, 
this still captures the majority of the SN population.

\begin{figure}[tb!]
\plotone{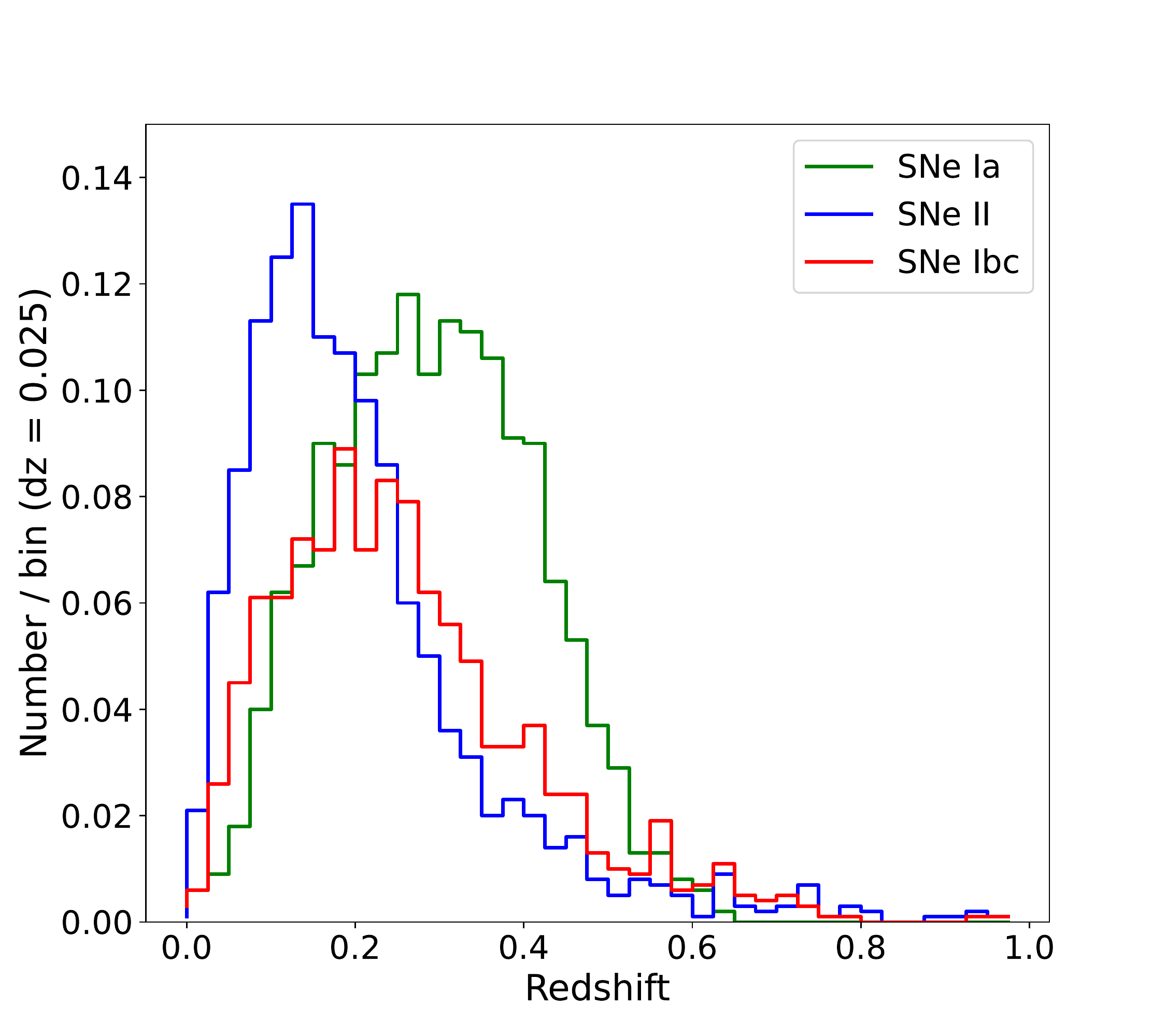}
\caption{Redshift distribution of SNe detected by the follow-up observations for the sky patch of $\Delta \Omega = 1 $ deg$^2$ with a 23~mag sensitivity limit. The total number of detected SNe is 1.5 (SNe Ia), 1.3 (SNe II), and 1.1 (SNe Ibc).
\label{fig:redshift_dist}}
\end{figure}
The number of Type Ia and II/Ibc SNe generated in the survey simulations is calculated according to the event rates from \citet{graur11} for Type Ia SNe, and from \citet{madau14}, 
proportional to the cosmic SFR, for CC SNe.
Among CC SNe, 70\% and 30\% of the total rates are assigned for Type II and Ibc SNe, 
respectively (see e.g., \citealt{graur17}).
If the simulated flux exceeds the observational sensitivity with $> 5 \sigma$ significance
at least once, we consider the object as detected.

The simulation identified $\approx$ 4 unrelated SNe in the localization area.  Figure \ref{fig:redshift_dist} shows the distribution of detected contaminants as a function of redshift. 
Because the detection sensitivity limit of 23 mag prevents 
distant transient sources from being detected (see Figure~\ref{fig:mag_redshift}), 
the number of contaminants is substantially reduced compared to the unbiased observation case.
Most of the detected contaminants are located beyond $z = 0.15$
as expected.
This is consistent with the analytic estimate calculated in Section \ref{subsec:statstic_followup}.
Note that 
the difference between Figures~\ref{fig:redshift_pdf} 
and \ref{fig:redshift_dist} arises from the different assumptions.
The former selects the closest object found in the unbiased
SNe sample, which is more appropriate for testing the chance coincidence 
background hypothesis, while the later case considers a realistic
magnitude-limited survey made with a medium-sized telescope.
%

\subsubsection{Discrimination of sources with small redshifts}
\label{subsubsec:discrimination}

It is ideal to perform real-time spectroscopy of all observed transients as it enables not only redshift measurements but also classification of the types of transients. For transients with $\lesssim 23$~mag, a typical exposure time of 1-2 hours is needed to obtain its redshift and transient type with 8-10~m class telescopes. Therefore, it would take 1-2 nights for all the discovered transients. 
A wide-field spectrograph with high multiplicity, such as the prime focus spectrograph on Subaru (\citealt{tamura16}) or MOONS on VLT (\citealt{cirasuolo20}) allows for the results to be obtained within a few hours.

However, these telescopes may not be available for observation at the given time. In this case, it is more practical to assign priorities to the follow-up observations based on the photometric redshift of the host galaxies.
Since the typical redshift range of the transient is $z < 0.6$, the photometric redshift given by the Pan-STARRS1 survey, covering the northern $3 \pi$ sky is sufficiently accurate ($\sim 3$\%) (\citealt{beck21}).
Photometric redshifts for the southern sky will also be available from the Vera~C.~Rubin observatory/LSST \citep{ivezic19}.
If the photometric redshift is $z < 0.15$, the transient is a strong candidate 
for the neutrino multiplet source, while if the host galaxy of a transient is $z > 0.15$, it can be regarded as a candidate for contamination. For the further identification of nearby neutrino source candidates and to study their nature, real-time spectroscopy as well as 
multicolor photometric observations are important, 
which will be discussed in the next sections.

\begin{figure}[tb!]
\plotone{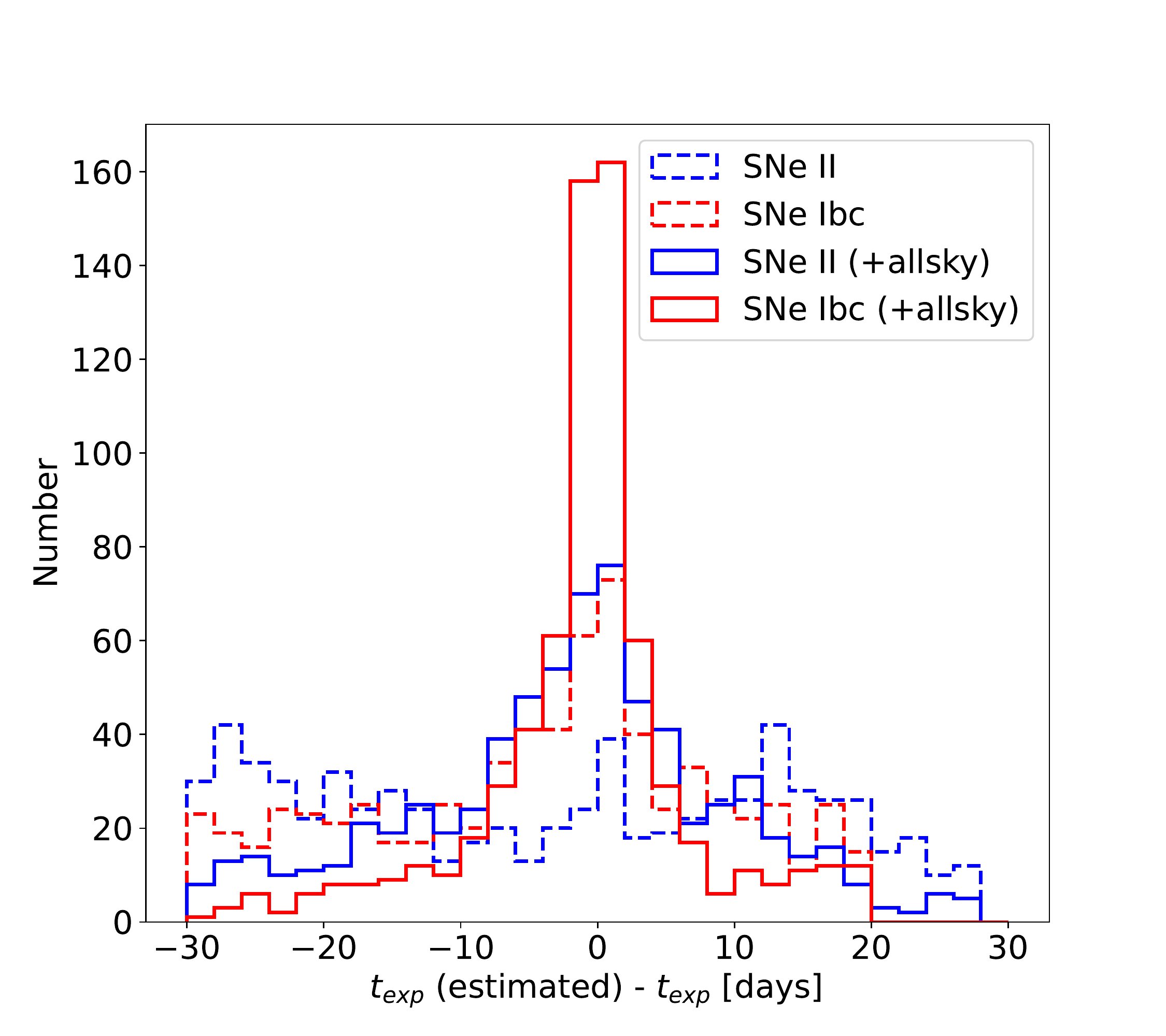}
\caption{Accuracy in the estimate of the explosion time. The dashed lines show the cases only with three epochs of observations, while the solid lines show the cases with additional all-sky data with a 20~mag depth.
\label{fig:deltat}}
\end{figure}

\subsubsection{Strategy to identify neutrino sources}
\label{subsubsec:identification}

The most likely redshift for the candidate of a neutrino multiplet source is $z \sim 0.03$ as indicated by the blue line in Figure \ref{fig:redshift_pdf}.
Figure~\ref{fig:mag_redshift} shows that the objects with such a redshift are expected to be brighter than 20-21~mag, and, hence, spectroscopic observations are feasible with a 2-4~m class telescopes.
Once the low redshift is confirmed, one can evaluate the $p$-value to test the statistical significance of the association, as discussed in Section \ref{subsec:statstic_followup}.

To further study the physics of the source, e.g., neutrino production mechanism and its timescale, it is also important to estimate the explosion time of the transient. Here, we demonstrate how accurately we can estimate the explosion time of the transients from the follow-up imaging observations discussed in Section~\ref{subsubsec:contaminants}.
We generate light curves of Type Ibc SNe and Type II SNe at $z = 0.0-0.15$, assuming they are the neutrino multiplet sources.
As a conservative case, the neutrino multiplet detection is assumed to happen 30 days after the explosion. This timescale in the SN evolution corresponds to the time duration of the interaction process between the SN ejecta and circumstellar material. We perform mock observations of sources for 10 epochs with a 5 day cadence, which assume continuous monitoring starting from the first search observations described in Section \ref{subsubsec:contaminants}.
 The first observation is assumed to start 1 day after the second neutrino detection, i.e., 31 days after the explosion.

Figure \ref{fig:deltat} shows the accuracy in the explosion date estimate by fitting the observed light curve with the template light curves.
The flat dashed lines indicate 
that the explosion time of the transient cannot be well determined.
This is because the observational data missed the rising and the peak of the light curves.
The accuracy for Type II SNe tends to be lower as their light curves are flat and featureless. If multiplet neutrino detection happens within 10 days for Type Ibc SNe, the estimate of the explosion time is accurate to within about 5 days as the observations can capture the rising phase, as demonstrated by, e.g., \citet{cowen10}.

All-sky time-domain data can improve these results.
The solid lines in Figure \ref{fig:deltat} show the same estimate but with 
all-sky data with three day cadence and 20~mag depth in the $r$-band from e.g. the Zwicky Transient Facility (\citealt{bellm19}). The accuracy is improved to $\sim 5-10$ days. As the multiplet candidates are located within the nearby universe, even relatively shallow data can improve the estimates of the explosion time if the cadence is sufficiently high.

Furthermore, multiwavelength data can be used to gain additional understanding. For the types of neutrino emission models involving shock interactions, which are expected in interacting SNe and TDE winds, gamma-ray, X-ray, and radio emissions are unavoidable ~\citep{Murase:2010cu,Murase:2020lnu}. Given that the sensitivities of these telescopes is sufficient, the detection of both thermal and nonthermal signals is likely for nearby sources. Gamma-ray, X-ray, and bright radio transients are less common than optical transients. This will help us better characterize the optical transients as true neutrino sources and examine the theoretical feasibility of neutrino-optical associations. 
For SNe, shock breakout emission from the stellar envelope or perhaps circumstellar material can reveal progenitor properties and constrain the explosion time estimation to hours depending on the progenitors.
Both real-time neutrino multiplet searches and multimessenger alerts are powerful tools for discovering potential sources of neutrinos. Such attempts include the Astrophysical Multimessenger Observatory Network (AMON) \citep{Smith:2012eu,AyalaSolares:2019iiy} and neutrino--gamma-ray coincident searches ~\citep{AMON:2019zxe,AMONTeam:2020otr}. 
This is also the case for TDEs. TDEs should happen in the nuclear region of a galaxy, and SNe far from the center can hence be readily removed. Both spectroscopic information and light curves will be needed for classifying transients in the nuclear region. Long-lived U-band, UV emission, and the Balmer line profiles are often seen in TDEs \citep{Stein:2020xhk,Reusch:2021ztx}. In addition, X-rays can be used as an additional probe. While TDE X-rays are more likely to be powered by the engine, interaction-powered SNe, including Type IIn can be accompanied by X-rays via shocks that also power the optical emission.

\section{Discussion} \label{sec:discussion}


Searches for multiple neutrino events by a $\sim$~1 km$^3$ neutrino detector such as IceCube and KM3Net enable us to probe the neutrino emission from sources with ${\mathcal E}_\nu^{\rm fl}\gtrsim 10^{50}$~erg with a flare timescale of $\lesssim 1$ month.
A null detection of neutrino multiplets at energies of $\gtrsim 50$~TeV would constrain the parameter space, $({\mathcal E}_\nu^{\rm fl}, R_0)$, of neutrino transient source models.
The region of the effective $N^{\rm M}_{\Delta \Omega} \gtrsim 10^{-6}$ 
in the right panel of Figure~\ref{fig:number_of_multiplet_s}
(see Equation~(\ref{eq:rate_of_multiplet_sources})) will be disfavored.
If TeV energy neutrino transients with a flare duration
of $\Delta T\sim 30$~days are indeed responsible for the major fraction of the neutrino diffuse cosmic background flux, this constraint 
under the condition of 
$E_\nu^2\Phi_\nu\approx 3\times 10^{-8}\ {\rm GeV}\ \ {\rm cm}^{-2}\ {\rm s}^{-1}\ {\rm sr}^{-1}$ constrains the neutrino source energy output and burst rate density as
\begin{equation}
    {\mathcal E}_\nu^{\rm fl} \lesssim  5\times 10^{51}\ {\rm erg}, \quad
    R_0  \gtrsim  2\times 10^{-8} ~{\rm Mpc}^{-3}~{\rm yr}^{-1},
    \label{eq:constraints_on_source_parameters}
\end{equation}
for $\xi_z\approx3$. 
This is consistent with the previous results~\citep{Esmaili:2018wnv,Murase:2016gly,Ackermann:2019ows} where $R_0\gtrsim 6\times{10}^{-8}~{\rm Mpc}^{-3}~{\rm yr}^{-1}~{(\xi_z/3)}^{-3}$.
This implies that rare sources, such as canonical high-luminosity GRBs and jetted TDEs, are already excluded from being the dominant sources~\citep[see also][]{Senno:2016bso,IceCube:2018omy}. Superluminous SNe are marginal, and may be critically constrained by near-future data.
Constraining the neutrino emission scenarios from more abundant sources, including LL GRBs, non-jetted TDEs, hypernovae and CC SNe, requires better detection sensitivities that can be achieved by IceCube-Gen2 \citep{IceCube-Gen2:2020qha} and KM3Net \citep{KM3Net:2016zxf}. 

There are some uncertainties in these constraints given our choice of parameterization in the construction of the generic neutrino emission models. Most notably, changing the value of $\alpha_\nu$, the power-law index of the neutrino spectrum, would lead to a systematic shift of $N^{\rm M}_{\Delta \Omega}$, the number of sources which produce multiple events. 
Our baseline is $\alpha_\nu=2.3$. However, softer spectrum scenarios, such as $\alpha_\nu\lesssim 2.6$, are still consistent with the IceCube's observations.
For example, $N^{\rm M}_{\Delta \Omega}$ is increased by a factor of $\sim 3$ if we assume $\alpha_\nu=2.6$. This is because more neutrinos are emitted at energies far below $\varepsilon_0=100$~TeV. On the other hand, it is statistically more consistent with the atmospheric neutrino background hypothesis to find such low-energy neutrinos. As shown in Figure~\ref{fig:p-values-doub}, it requires $E_\nu\gtrsim $~30 TeV in order to claim the statistically meaningful  association with a transient radiation from a source. As a result, the effective p-value calculated by Eq.~(\ref{eq:iceCube_multiplet_sensitivity}) does not significantly depend on the assumption of $\alpha_\nu$ for a given set of the source parameters (${\mathcal E}_\nu^{\rm fl}$ and $R_0$). 
If the atmospheric background neutrino rate remains the same, the sensitivity (constraints) presented in the middle (right) panel of Figure~\ref{fig:number_of_multiplet_s} is, thus, still valid.

We have been discussing the transient flare timescale, $\Delta T$, of $\sim 30$
days so far. It is possible that some optical transients are shorter in duration.
The number of sources to yield multiplet sources, $N^{\rm M}_{\Delta\Omega}$, 
is unchanged for different $\Delta T$, as long as 
the multiplet search time window $T_{\rm w}$ is longer than $\Delta T$ to monitor
the entire neutrino flare phenomena. However, the effective number of sources 
required to reject the atmospheric neutrino background hypothesis is increased
for a flare with a time scale shorter
than 30 days. For example,  $N^{\rm M}_{\Delta\Omega}$  would be increased 
by a factor of 2 for $T_{\rm w}=10$~days.
The resultant constraints on 
${\mathcal E}_\nu^{\rm fl}$ and $R_0$ as represented by
Eq.~(\ref{eq:constraints_on_source_parameters}) can be more stringent by approximately
 a factor of 4. 
Our choice throughout this paper to characterize the longer timescale search therefore represents a conservative estimate.
Any transient much longer than $\Delta T$ of 30 days,
however, can hardly yield a detectable multiplet because the expected number of atmospheric
neutrinos during the flare time frame becomes $\mu_{\rm atm}\gtrsim 1$.

The other major uncertainty in our generic neutrino emission models
arises from the cosmological source evolution $\psi(z)$.
We assumed it to trace the SFR as represented by Eq.~(\ref{eq:sfd}).
Departure from an SFR-like evolution hardly changes $N^{\rm M}_{\Delta\Omega}$
as the multiplet sources are mostly confined within the nearby local universe.
The evolution factor is rather sensitive to the intensity of the diffuse
cosmic background radiation, as shown in Eq.~(\ref{eq:evolution_scaling}).
For object classes with no evolution, $\xi_z\approx 0.6$, which results
in $\sim$ 20 \% of the flux in the case when it follows an SFR-like evolution.
As a result, the constraints on ${\mathcal E}_\nu^{\rm fl}$ and $R_0$
by the null detection of the multiplet events become more stringent as
\begin{equation}
    {\mathcal E}_\nu^{\rm fl} \lesssim  2\times 10^{50}\ {\rm erg}, \quad
    R_0  \gtrsim  3\times 10^{-6} ~{\rm Mpc}^{-3}~{\rm yr}^{-1}.
    \label{eq:constraints_on_source_parameters_without_evolution}
\end{equation}
This is because neutrino sources with weaker evolution must be more populous
in the local universe to reach the observed diffuse cosmic background flux,
and the neutrino multiplet event searches are sensitive to the emission
from nearby sources. As some of the transient source candidates may be
likely to be only weakly evolved ({\it e.g.,} TDEs), the sensitivities
assuming the SFR-like evolution as the baseline in our study are again conservative.

Increasing the precision of the neutrino localization is a plausible way to reduce the atmospheric background neutrino rate and further improve sensitivity. We considered $\Delta\Omega =1\ {\rm deg}^2$ which is consistent with the present angular resolution of muon tracks measured by IceCube. Since the atmospheric background rate, $\mu_{\rm atm}$, is proportional to $\Delta\Omega$, reducing $\Delta\Omega$ would result in a significant improvement in the sensitivity, as shown by Eq.~(\ref{eq:extended_likelihood_limit}) where $\mathcal{L}^{\rm BG}\propto \mu_{\rm atm}^2$.
As a deep water neutrino telescope such as KM3Net expects better 
than $1~{\rm deg}^2$ localization, it will cover more parameter space
on (${\mathcal E}_\nu^{\rm fl}, R_0$). It should also be remarked that
the number of contaminants in optical followup observations should be substantially
reduced in this case. A neutrino doublet detection with $0.25~{\rm deg}^2$
localization error would expect only a single contaminant by a followup observation
with the 23 mag sensitivity limit, which may realize a contamination-free 
optical counterpart search with the photometric redshift information.

In the context of the angular resolution factor, one can add the pdf of angular distance from a point source location to the likelihood construction given by Eq.~(\ref{eq:extended_likelihood_final}) in order to enhance sensitivity further. 
The angular pdf is often described by a Gaussian function with $\sigma$ consistent with the angular resolution, for example, in the point source emission searches by IceCube.
The pdf depends highly on 
the reconstruction quality in each of the events measured by a neutrino detector. Its implementation into the likelihood construction is beyond the scope of this paper. The sensitivity shown in Figure~\ref{fig:number_of_multiplet_s}
is a conservative estimate.

Another simplification introduced in our study is the assumption that the energy of a neutrino event $E_\nu$ is known without any error. However, errors in estimating the neutrino energy cannot be avoided in observations. In order to assess this impact on the sensitivity, we convolved the energy pdf $P^{\rm E}(E_\nu)$ with the Gaussian error function, with $\sigma_{\rm log E}=0.2$, in the likelihood construction in Eq.~(\ref{eq:extended_likelihood_final}).
The p-values in this case are shown by the dashed curves in Figure~\ref{fig:p-values-doub}. The energy threshold to claim the multiplet source association becomes higher by $\sim 40\%$, which may result in $\sim 35$~\% degradation of the effective p-value to support the signal hypothesis displayed in the middle and right panels of Figure~\ref{fig:number_of_multiplet_s}.

The statistical significance of the neutrino source identification using the multiplet detection
somewhat depends on the local SNe number density, which is assumed to trance the $B$-band luminosity density
in our modeling (see Table~\ref{tab:p-value_tbl}).
It has been pointed out that the SN rate per galaxy $B$-band luminosity is 
known to show a dependence on the galaxy type~\citep{Li:2010kd} and
the SN density per galaxy is not exactly constant. However,  our approach in this paper 
uses the total SN rate across various types of galaxies. This is more robust as it is estimated with a large number of galaxies in a 1 deg$^2$ patch of the sky and the galaxy-type-dependent variation is diminished.

\section{Summary} \label{sec:summary}
Global searches for multiple neutrino events within a time window of $T_{\rm w}\lesssim 30$~days in the TeV-PeV energy neutrino sky with a $\sim 1$~km$^3$ neutrino telescope allow for the study of neutrino emission from long-duration sources with $\textstyle{{\mathcal E}_\nu^{\rm fl}\gtrsim 10^{50}}$~erg and $\textstyle{R_0\lesssim 3\times10^{-6}~{\rm Mpc}^{-3}~{\rm yr}^{-1}}$ (the domain of $P_{\rm m}^{\rm eff}\gtrsim 10^{-6}$ in the middle panel of Figure~\ref{fig:number_of_multiplet_s}). This covers the parameter space including neutrino transient emission from SL SNe, LL GRBs, and non-jetted TDEs. 
Therefore, the absence of
neutrino multiplet detections with the current generation
of detectors can constrain models involving these sources.
However, for the full exclusion of the regions of the relevant parameter space, future larger neutrino
detectors, such as IceCube-Gen2, are needed.
Requiring multiple neutrino detections limits the distance to possible neutrino emitters, which results in only a few contaminants, even in the extremely populated optical transient sky containing various types of SNe.
Redshift measurements of each of the possible counterparts brighter than 23rd magnitude can tell whether a given counterpart is likely to be associated with the neutrino multiplet detection. For example, finding an SN-like transient at $z=0.04$ in an optical followup observation leads to $\sim 2.7\sigma$ significance against the chance coincidence background hypothesis
(See Table~\ref{tab:p-value_tbl}). 
This demonstrates that obtaining multimessenger observations triggered by a neutrino multiplet detection is a practically feasible approach to identifying TeV-energy neutrino sources, for which hidden neutrino sources may be dominant, and study their emission mechanism
as well as particle acceleration in dense environments.

\begin{acknowledgments}
The authors are grateful to Yousuke Utsumi for his useful suggestions on the pilot study
based on the GLADE galaxy catalog. We also thank Brian Clark for his careful reading 
of the manuscript.
This work by S.Y.,~A.I.,~and N.S is 
supported by JSPS KAKENHI Grant No.~18H05206, 18H05538,
and Institute for Global Prominent
Research (IGPR) of Chiba University.
K.M. is supported by the NSF Grant No.~AST-1908689, No.~AST-2108466 and No.~AST-2108467, and KAKENHI No.~20H01901 and No.~20H05852.
M.T. is supported by JSPS KAKENHI Grant No.~17H06363, No.~19H00694, No.~20H00158, and No.~21H04997.
\end{acknowledgments}

%

\vspace{5mm}
\facilities{IceCube, Subaru, Rubin, Blanco, ZTF}
\software{sncosmo \citep{barbary16}}


\bibliography{reference}{}
\bibliographystyle{aasjournal}



\end{document}